\documentclass[preprint, 12pt]{elsarticle}

\pdfoutput=1

\usepackage{lineno,hyperref}
\modulolinenumbers[100]

\usepackage{multirow}
\usepackage{hhline}

\usepackage[english]{babel}
\usepackage[utf8]{inputenc}
\usepackage{algorithm}
\usepackage[noend]{algpseudocode}

\usepackage{color}

\usepackage{array}

\usepackage{graphicx}
\usepackage{amsmath}
\usepackage{amssymb}
  \usepackage[caption=false,font=footnotesize]{subfig}
\journal{Journal of Computer Networks}

\begin{document}

\begin{frontmatter}

\title{Towards Limited Scale-free Topology with Dynamic Peer Participation}
%\tnotetext[mytitlenote]{Fully documented templates are available in the elsarticle package on \href{http://www.ctan.org/tex-archive/macros/latex/contrib/elsarticle}{CTAN}.}

%% Group authors per affiliation:
%\author{Xiaoyan Lu\fnref{myfootnote}}
%\address{Department of Computer Science, Rensselaer Polytechnic Institute}
%\fntext[myfootnote]{Since 1880.}

%% or include affiliations in footnotes:
\author[rpi]{Xiaoyan Lu}
\ead{lux5@rpi.edu}

\author[vcu]{Eyuphan Bulut\corref{cor1}}
\cortext[cor1]{Corresponding author}
\ead{ebulut@vcu.edu}

\author[rpi,poland]{Boleslaw Szymanski}
\ead{szymab@rpi.edu}

\address[rpi]{Department of Computer Science, Rensselaer Polytechnic Institute, Troy, NY, 12180}
\address[vcu]{Department of Computer Science, Virginia Commonwealth University,\\ Richmond, VA, 23284}
\address[poland]{Wroclaw University of Technology, 50-370 Wroclaw, Poland}

\begin{abstract}
Growth models have been proposed for constructing the scale-free overlay topology to improve the performance of unstructured peer-to-peer (P2P) networks. However, previous growth models are able to maintain the limited scale-free topology when nodes only join but do not leave the network; the case of nodes leaving the network while preserving a precise scaling parameter is not included in the solution. Thus, the full dynamic of node participation, inherent in P2P networks, is not considered in these models. In order to handle both nodes joining and leaving the network, we propose a robust growth model E-SRA, which is capable of producing the perfect limited scale-free overlay topology with user-defined scaling parameter and hard cut-off. Scalability of our approach is ensured since no global information is required to add or remove a node. E-SRA is also tolerant to individual node failure caused by errors or attacks. Simulations have shown that E-SRA outperforms other growth models by producing topologies with high adherence to the desired scale-free property. Search algorithms, including flooding and normalized flooding, achieve higher efficiency over the topologies produced by E-SRA.
\end{abstract}

\begin{keyword}
Scale-free networks; Peer-to-Peer networks; Scalable; Overlay networks.
\end{keyword}

\end{frontmatter}

\linenumbers

\section{Introduction}

{I}{n} addition to the specific search strategies and resource allocation methods, the overlay topologies (i.e. logical connectivity graph) have significant impact on the performance of unstructured peer-to-peer (P2P) networks. It has been shown that the scale-free topology with the power-law distribution of node degrees is by definition well suited for P2P networks~\cite{scalefree:hui}. This is because such topology has a logarithmically scaled diameter~\cite{scalefree:diameter} (ranging from $O(\ln N) $ to $O(\ln \ln N)$), is highly tolerant to random failures~\cite{yuksel:y1} and network congestion~\cite{yuksel:y3}, and it also is highly synchronizable~\cite{yuksel:y2}.

In the original Barabasi-Albert (BA) model~\cite{scalefree:BA}, when a new node joins the network, its likelihood to connect to a node increases with this node's degree. This behavior, also known as "Preferential Attachment", generates a network with scale-free features. The BA model, however, does not consider some important features of P2P applications, including the hard cut-off on degree and dynamic peer participation~\cite{p2p:churn}. The hard cut-off, which restricts the feasible topologies to limited scale-free networks, is required because users in P2P networks usually are not willing to serve as hubs (i.e. nodes with high degree) because of the high bandwidth required to serve the ensuing traffic. Moreover, users usually keep joining and leaving the network periodically, which affects the degree distribution in an unpredictable manner. Efforts have been made to build the limited scale-free overlay topology for P2P networks. Previous studies ~\cite{scalefree:yuksel,scalefree:grow,scalefree:Gaian} have proposed cost-efficient growth models to construct the limited scale-free topologies with hard cut-offs with nodes constantly joining the network. However, the dynamic of nodes' removal~\cite{p2p:churn} is another inherent property of P2P networks and it is not taken into account in these growth models. In previous studies ~\cite{yuksel:adhoc, yuksel:popularity}, although nodes are allowed to leave, a precise scaling parameter can not be produced as specified.

%churn nature of p2p network, removing nodes
Scale-free topologies are relatively robust in face of the removal of randomly chosen nodes but are very vulnerable to the removal of hubs~\cite{scalefree:hubremove,scalefree:hubremove2}. In P2P networks, a selfish hub may quit and rejoin the network to avoid high communication costs, potentially distorting the degree distribution of the topologies. Admittedly, the degree distribution of a scale-free topology is not significantly influenced by infrequent node removal. However, when node removals are significant fraction of node additions over some period, the accumulated effects will eventually destroy the scale-free topology. In order to preserve the scale-free topology, a mechanism is needed to keep the power-law degree distribution regardless of the frequency of node removals.

One challenge to preserve a scale-free topology is to avoid using global knowledge because the communication cost required to obtain it grows linearly with the size of network. Yet, degree distribution is a global property. In addition, when a node is removed, each of its neighbors has one connection terminated. Thus, scale-free feature must be controlled with a cost-efficient approach which does not require global information. On the other hand, it is important to preserve an exact scaling parameter for a power-law scale-free topology because the best performance of particular P2P application is achieved when a specific scaling parameter is chosen. But previous decentralized protocols can only produce overlays with an inexact and constant scaling parameter, depending on the size of the network.

\iffalse
This scenario is analogous to maintaining a hierarchical management structure in an organization. In order to keep the original hierarchical management structure without hiring new employees, if many middle managers resign, some of the low-level managers must be promoted while some of the top managers have to be downgraded.
\fi

Here, we introduce the \textbf{E}nhanced \textbf{S}emi-\textbf{R}andomized Growth \textbf{A}lgorithm (E-SRA), which preserves the power-law degree distribution in an overlay topology with nodes dynamically joining and leaving the network. It allows arbitrary nodes to be removed from a network and does not require collecting global information about the topology. Our approach assumes only local information is available, i.e. degree of the neighbors of the removed node. One advantage of our approach is that it provides partial tolerance to failing nodes. If a single node\footnote{We assume the nodes crash incrementally. When a group of connected nodes crash at the same time, some global restoration algorithm would be more appropriate.} fails due to attacks or errors, the neighbors are able to detect the failure and work to preserve the power-law distribution. In terms of the message complexity, one broadcast is needed to remove one node. If nodes randomly leave the network, the average number of the point-to-point messages sent by a single node to preserve the power-law degree distribution is linear to the maximum (also called hard cut-off) degree of the nodes. Combined with the growth model proposed in previous work~\cite{scalefree:grow}, scale-free overlay networks can be maintained efficiently with nodes freely joining and leaving. As discussed in \cite{scalefree:grow}, it is important to adhere to exact scaling parameters because only then full advantage of properties such as slowly growing diameter hold. Moreover, capability to control post-construction parameters gives the flexibility to have a desired $\gamma$ that will give the best efficiency in the search algorithm in use.

%contributions
 Simulations have shown that our approach can generate the perfect limited scale-free topologies with different patterns of adding/removing nodes while other models presented in the literature ~\cite{scalefree:yuksel,scalefree:grow,scalefree:Gaian} have failed to handle node removal, especially when nodes with high degrees leave the network frequently. Moreover, the overlay topologies constructed by E-SRA provide better search performance in various settings; here, we considered search algorithms including flooding, where messages are forwarded to all neighbors, and normalized flooding, where messages are forwarded to $k$ (i.e. the minimum degree) randomly selected neighbors. 
 
 The major properties of E-SRA include the following:
 \begin{itemize}
     \item Tolerance to dynamic peer participation: Nodes with any degree are free to leave and join the network while the power-law distribution is preserved.
     \item Partial tolerance to failures: the topology can recover from a single node failure as long as the neighbors of the failing node are alive at the moment of the failure. 
     \item Scalability: no global information is required to add or remove a node.
     \item Flexible parameter settings: a topology with user-defined parameters (i.e. the minimum/maximum degree, scaling parameter $\gamma$) is created to ensure its optimal performance in applications such as search algorithms.
 \end{itemize}
 
%section
Section \ref{sec:related} introduces the previous work on growth models for limited scale-free topologies. Section \ref{sec:aa} presents the detailed algorithms and its analysis. Section \ref{sec:sim} describes simulation results. The discussion is presented in Section \ref{sec:discuss} and the conclusions are included in Section \ref{sec:conclusion}.

\section{Related Work}\label{sec:related}
The scale-free property is shown to exist in many natural or artificial systems, such as protein-protein interaction networks\cite{scalefree:protein}, the Internet\cite{s:1}, the World Wide Web\cite{s:2}, and scientific collaboration networks\cite{s:3}. The degree distribution in these networks follows the power-law: $P(i) \sim  i^{-\gamma}$, where $P(i)$ is the fraction of nodes with degree $i$ and $\gamma$ is the scaling parameter which varies between different types of networks ($2 \leq \gamma \leq 3$ in most cases). In a limited scale-free topology, only nodes with degrees smaller than the hard cut-off (i.e. the maximum) degree have degree distribution that follows the power-law.

The scale-free topology has some good properties, including high tolerance to random attacks~\cite{yuksel:y1}, high synchronizability \cite{yuksel:y2} and resistance to congestion \cite{yuksel:y3}. For this reason, several growth models are proposed to construct the scale-free overlay topology. The BA model \cite{scalefree:BA} manages to explain the evolution of scale-free topologies by a core principle named "Preferential Attachment". But it is not practical in real P2P applications because the global information is required to maintain it. To address this issue, HAPA \cite{scalefree:yuksel}, Gaian \cite{scalefree:Gaian}, subPA \cite{yuksel:adhoc} and SRA \cite{scalefree:grow} algorithms were introduced to construct the scale-free overlay topology with partial or no global information.

"Preferential Attachment" \cite{scalefree:BA} means a new node is more likely to connect to heavily linked nodes when it joins the network. The BA model has some disadvantages as the growth model for the overlay topology of P2P networks. Firstly, it does not provide hard cut-offs. Since a heavily linked node uses a lot of bandwidth in P2P networks, nodes usually are not willing to maintain high degrees. For this reason, a user-defined hard cut-off (i.e. the maximum) degree is imposed lower than the natural cut-off arising in the BA model. The imposed hard cut-offs restrict the feasible overlays to the limited scale-free topologies, which are more practical. Moreover, the BA model also requires the global information about the topology to add connections when a new node joins. In real-world applications, however, the communication cost of obtaining the global information is prohibitive. Therefore, a distributed approach that constructs the topology without global information is desired. 

In~\cite{scalefree:yuksel}, authors study the construction of limited scale-free overlay topologies for unstructured peer-to-peer networks. In the Hop-and-Attempt Preferential Attachment (HAPA) algorithm~\cite{scalefree:yuksel}, a new node joining the network connects to $k$ (i.e. the minimum degree) nodes in a random route starting from a randomly chosen node. This scheme works because high degree nodes are more likely to occur in a random route than nodes with low degree. The hard cut-off is used to avoid "superhubs", which are nodes with degrees linear to the network size. HAPA algorithm produces the topologies with degree distribution approximately following the power-law with scaling parameter $\gamma=3$. 

Gaian~\cite{scalefree:Gaian} algorithm is proposed for distributed database systems whose efficiency and reliability depends on the overlay topology. In this algorithm, a new node broadcasts a message when it joins the network using computing with time principle~\cite{yeye:time}. Each receiver computes the maximum time of delay, $t_v$, which is proportional to the inverse of its degree, and chooses the time of delay $t_d$ in the interval $[0, t_v]$ uniformly randomly. Instead of replying to the sender instantly, the receiver waits the time of delay $t_d$ and then replies. In this way, nodes with higher degrees are likely to wait shorter period. The new node connects to the first $k$ responders. It gives a better chance to the new node to connect to nodes with high degrees. This mechanism, which reduces the communication overhead by allowing nodes to self-select themselves according to their fitness to the desired property, is also known as computing with time~\cite{yeye:time}. The communication cost for selection is constant in the number of candidates. Similar to HAPA algorithm, Gaian algorithm cannot produce an overlay topology with the user-defined scaling parameter.

    Although the growth models such as Gaian and HAPA can construct a limited scale-free topology efficiently and are also easy to implement, post-construction parameters of network structures (i.e. the scaling parameter related to the search performance) cannot be adjusted. In \cite{scalefree:grow}, the authors propose a flexible growth model that can produce a limited scale-free topology with user-defined parameters. The \textbf{S}emi-\textbf{R}andomized Growth \textbf{A}lgorithm (SRA)\cite{scalefree:grow} requires no global information and imposes a hard cut-off on degrees. In SRA, when a node joins the network, it broadcasts a message containing the desired degrees of the $k$ new neighbors. These degrees are computed according to the given network parameters. The receivers with the desired degrees reply to the new node using computing with time rule~\cite{yeye:time}. The new node connects to the first $k$ responders. The scaling parameter, as well as the hard cut-off, can be defined by users in advance. So, the SRA model is able to produce overlay topologies, over which the efficiency of applications such as search algorithms is maximized. The constraints of feasible values of these network parameters are presented in \cite{yeye:perfect}. SRA outperforms other growth models by producing overlay topologies with perfect matching to the arbitrary power-law degree distribution.

% MEntion here Yuksel's ad hoc and popularity papers
In \cite{yuksel:adhoc}, the authors propose an ad-hoc limited scale-free network model, subPA\footnote{The approach proposed in \cite{yuksel:adhoc} is named after "subPA" because it performs the Preferential Attachment (\textbf{PA}) process in a chosen \textbf{sub}-graph. Originally, the authors of \cite{yuksel:adhoc} did not use any abbreviation for their approach.}, which performs the \textbf{P}referential \textbf{A}ttachment process in a chosen \textbf{sub}set of nodes. When a new node joins the network, it constructs a set of nodes reachable in at most $\tau_j$ hops from a random existing node and connects to $k$ nodes in this set. The probability a node being connected is proportional to its degree. When a node is removed, its neighbors construct a set of nodes reachable in $\tau_l$ or less hops from the deleted node, and each neighbor connects to one node using the similar preferential attachment rule. The hard degree cut-off is achieved by prohibiting nodes with maximum degree from accepting new connections. It is worth noting that this model becomes the preferential attachment with global information when $\tau_j$ value is large and $\tau_l$ is zero and a BA network with $\gamma = 3$ is obtained. In \cite{yuksel:popularity}, the subPA model is extended to consider the popularity of a node, which is defined as the summation of the popularity of the items it holds. The likelihood that an existing node being connected is proportional to the linear combination of its degree and popularity. In this way, the degree of the nodes which hold popular downloadable items increases, leading to reduced query response time. But this framework could not specify the scaling parameter of the produced topologies and it does not consider the removal patterns which are likely to destroy the scale-free property, such as frequent removal of hubs. The features of HAPA, Gaian, subPA and SRA algorithms are compared in Table \ref{table:compare}.

To the best of our knowledge, however, allowing nodes to leave network during overlay maintenance while producing the precise scaling parameter has not been studied. Yet, in peer-to-peer networks, nodes are likely to join and leave the network frequently. With such dynamic peer participation, the scale-free topology produced by previous growth models are affected by node removal, especially when hubs are removed. This effect can accumulate, negatively impacting the performance of P2P networks, which are built on top of the overlays. In this paper, we propose a robust model, in which nodes are allowed to leave the network in an arbitrary pattern. Combined with the growth model proposed in our previous work \cite{scalefree:grow}, the scale-free topology is maintained while nodes are allowed to freely join and leave the network.

\iffalse
There are two main measurements for these growth models. First, limited scale-free network topology, which has a hard cut-off on degree, is constructed. In real networks, single node often sets a limitation on the number of connections it has. In peer-to-peer networks, users usually are not willing to afford high bandwidth in the network. The second principle is low communication overhead. As mentioned above, communication cost is much higher if global information is required to determine which peer to connect. When nodes join and quit frequently, a consistent global topology information is very difficult to be maintained at every node. Thus, distributed algorithm using local information is preferred in the construction of scale-free topology.
\fi

Other decentralized or self-organized algorithms which construct the overlay topologies in a wide variety of applications for unstructured peer-to-peer systems include \cite{suggest:Montresor04arobust,suggest:snyder2009myconet,suggest:dumitrescu2012clustering,suggest:liu2008erasp}. These protocols construct overlay topologies efficiently based on the superpeer which operates as the server for a set of clients. They trade the strict adherence to the precise degree distributions for increased resilience in the face of frequent and simultaneous join and leave events. Compared with these protocols, the primary concern of our work is to construct power-law topologies with a precise and controllable scaling parameter so that P2P applications achieve the best performance over them. And our work does not assume the existence of such superpeers for maintaining overlay topologies, which in purely P2P applications are not assumed to exist.

\section{Approach and Analysis} \label{sec:aa}
We describe the problem and propose our approach which focuses on preserving the power-law degree distribution while nodes dynamically join and leave the network. Below, we discuss leaving only since the joining part is presented in detail in \cite{scalefree:grow}.

\subsection{Problem Formulation} \label{sec:problem}
The degree of a node is defined as the number of connections it has in an overlay topology. Using the same notation as in \cite{scalefree:grow}, the fraction of nodes with degree $i$ is denoted as $P_i$,
\begin{equation}
P_i = \frac{N_i}{N}
\end{equation}
where $N_i$ is the number of nodes with degree $i$ and $N$ is the total number of nodes. 

In a scale-free topology, degree distribution follows the power-law: $P_i \sim i^{-\gamma}$ where $\gamma$ is a constant. In a limited scale-free topology, $P_i$ follows the power-law for $i<m$, where $m$ is the hard cut-off (i.e. the maximum) degree.

The value of $P_i$ in a limited scale-free topology is given by Eq. [7] in \cite{scalefree:grow} as $f_i$ as a function of the maximum degree $m$, the minimum degree $k$ and the scaling parameter $\gamma$, 
\begin{equation} \label{eq:fi1}
f_i = \frac{m - 2 k}{ i^\gamma \sum_{j=k}^{m-1} \frac{m-j}{j^\gamma}}  \quad \text{for} \quad i < m
\end{equation}
and $f_m$, that does not need to follow the power-law distribution, is given as,
\begin{equation} \label{eq:fi2}
f_m = 1 - \sum_{i=k}^{m-1} f_i
\end{equation}
The goal is to maintain the degree distribution as $f_i$ for $i=k,\ldots,m$ while nodes with arbitrary degrees are added or removed\footnote{We are interested in preserving the power-law distribution because of the performance improvements it brings. Our approach actually solves the general problem of keeping any fixed degree distribution.}. For the sake of simplicity, we assume one node is removed at a time and the node to be removed is denoted as node $R$ and the number of its neighbors is denoted by $b$.

\subsection{Atomic Operations}
In order to preserve the power-law distribution, we made three observations about the dynamics of the topology when a single node $R$ is removed.\\
  \textbf{Observation 1}: If an arbitrary node $R$ is removed, the neighbors of $R$ are able to detect the removal because each of them has one connection terminated.\\
  \textbf{Observation 2}: If an arbitrary node $R$ is removed, the degree of each neighbor of $R$ decreases by one, influencing the resulting degree distribution. One simple countermeasure is to connect each neighbor with a random node so that the degrees of $R$'s neighbors stay the same.\\
  \textbf{Observation 3}: If an arbitrary node $R$ with degree $b$ is removed, the numbers of nodes with degrees both lower and higher than $b$ should change to preserve the power-law distribution. This scenario is analogous to maintaining hierarchical management structure in a large company. If many middle managers resign, in order to keep the original ratio of managers on different levels without hiring new employees, some low-level managers must be promoted while some managers have to be downgraded.

The main challenge here is that the pattern how the node degrees change is unpredictable. In order to handle the dynamics, each neighbor of $R$ can connect to a new node, then all neighbors of $R$ will have the same degrees. It means there are a total of $b$ connections added by the neighbors of $R$, and the degrees of the nodes on the other end of these $b$ connections will increase by one. If the nodes on the other end of these $b$ connections are chosen carefully, then the change of degrees is predictable.

In addition, from \textit{Observation 3}, we know that in some cases, certain connections should be removed to preserve the power-law degree distribution. How to locate these connections? A solution that we use here is based on the following observation. Consider a node $X$ that is a neighbor of node $R$ and two nodes $A$ and $B$ that are not but they are connected to each other. When node $A$ connects to $X$ and also terminates its connection to $B$, then, nodes $A$ and $X$ keep their original degrees while the degree of node $B$ decreases by one. If node $B$ is carefully chosen to have the right degree, then the degrees are changed as desired.

To sum up this idea, there are two types of operations that can be conducted by each neighbor of $R$:
\begin{itemize}
    \item PUSH: The neighbor of $R$ connects to a new node $A$.
    \item SHUFFLE: Besides connecting to a new node, the neighbor of $R$ also asks the new node $A$ to terminate one connection to some node $B$.
\end{itemize}

One PUSH will increase the degree of node $A$ by one and one SHUFFLE will decrease the degree of node $B$ by one. In the rest of this paper, we say a neighbor of $R$ PUSHes on degree $i$, if it connects to a new node $A$ with degree $i$, causing $A$ to increase its degree to $(i+1)$. And we say a neighbor of $R$ SHUFFLEs on degree $i$, if it connects to a new node $A$ and asks $A$ to terminate its existing connection to a node $B$ of degree $i$, causing $B$ to decrease its degree to $(i-1)$. The neighbor of $R$ will keep the original degree no matter whether it PUSHes or SHUFFLEs.

How many PUSHes and SHUFFLEs should be assigned to the neighbors of $R$? Since the degree distribution of a limited scale-free topology does not change, the same average degree $2k$ should be preserved. Thus, if one node is removed, totally $k$ connections should be removed. So, when $b$ connections of node $R$ are removed, $(b-k)$ connections should be added to keep the average degree at $2k$. This could be achieved by $(b-k)$ PUSHes and $k$ SHUFFLEs per removal because each PUSH adds one connection and each SHUFFLE does not change the total number of connections.
 
 \subsection{Analysis} \label{sec:analysis}
We are interested in the degree of every PUSH and SHUFFLE when a single node is removed. Let $D_i$ denote the number of SHUFFLEs on degree $i$ and $I_i$ denote the number of PUSHes on degree $i$. Since $(b-k)$ PUSHes and $k$ SHUFFLEs are needed, we have,
\begin{align} \label{eq:sumIsumD}
 \sum_{i=k}^{m-1} I_i &= b-k\\
 \sum_{i=k+1}^{m} D_i &= k
\end{align}
These SHUFFLEs and PUSHs make $D_i$ nodes \textit{decreasing} their degree from $i$ to $(i-1)$ and $I_i$ nodes \textit{increasing} their degree from $i$ to $(i+1)$. Also, $I_m = 0$, $D_k = 0$ because the degrees of all nodes are kept in range $[k, m]$. $I_i$, $D_i$ are non-negative for $i\in[k,m]$.

Let's consider the total number of nodes with degree $k$ after node $R$ is removed from a network of size $n$. Before removal, there were $f_k n$ nodes originally of degree $k$. Additional $D_{k+1}$ nodes originally with degree $(k+1)$ are added and $I_{k}$ nodes are moved from this count by SHUFFLEs and PUSHes. If the fraction of nodes with degree $k$ is still $f_k$, we have,
\begin{equation} \label{eq:0}
f_k (n-1) = f_k n - I_k + D_{k+1} 
\end{equation}
where $n$ is the total number of nodes before $R$ quits. Similarly, the degrees of $I_{d}$ nodes increase from $d$ to $(d+1)$. The degrees of $I_{d-1}$ nodes increase from $(d-1)$ to $d$. The degrees of $D_{d+1}$ nodes decrease from $(d+1)$ to $d$. And the degrees of $D_d$ nodes decrease from $d$ to $(d-1)$. If the fraction of nodes with degree $d$ remains $f_d$, then
\begin{equation} \label{eq:1}
 f_d (n-1) = f_d n - I_d + I_{d-1} + D_{d+1} - D_d 
\end{equation}
for $k < d < m$ and $d \neq b$. Since node $R$ itself is removed,
\begin{equation} \label{eq:2}
f_b (n-1) = f_b n - I_d + I_{d-1} + D_{d+1} - D_d - 1 
\end{equation}
Due to the hard cut-off, nodes with degree $m$ should not accept any new connections,
\begin{equation} \label{eq:3}
f_m (n-1) = f_m n + I_{m-1} - D_m 
\end{equation}
Simplifying Eqs [\ref{eq:0},\ref{eq:1},\ref{eq:2},\ref{eq:3}], we obtain that for $k \leq i \leq b-1$,
\begin{equation} \label{eq:12}
I_{i} - D_{i+1} = \sum_{j=k}^{i} f_j
\end{equation}
and for $  b \leq i < m$,
\begin{equation} \label{eq:13}
I_{i} - D_{i+1} = \sum_{j=k}^{i} f_j - 1
\end{equation}
Let non-negative vectors $\vec{D} = (D_{k+1}, D_{k+2}, \ldots, D_{m})^T$, and $\vec{I} = (I_{k}, I_{k}, \ldots, I_{m-1})^T$ be such that,
\begin{align} \label{eq:vec}
    \vec{I} - \vec{D} &= \begin{bmatrix}
           f_k \\
           f_k+f_{k+1} \\
           \vdots \\
           \sum_{i=k}^{b-1} f_i \\
           \sum_{i=k}^{b} f_i - 1 \\
           \vdots \\
           \sum_{i=k}^{m-1} f_i - 1 \\
         \end{bmatrix}
\end{align}
with the $L^1$ norm $\|\vec{I}\|_1 = b - k$, $\| \vec{D} \|_1 = k$. The solution to Eq. [\ref{eq:vec}] depends on the degree distribution $f_i$, the degree of removed node $b$ and the hard cut-off $m$, but is independent from the current network size $n$. It allows us to design an algorithm that does not need any global information.\\
One simple solution to Eq. [\ref{eq:vec}] is,
\begin{align} \label{eq:sol1}
I_i^* &=  \begin{cases}
1 &\text{$ k \leq i < b$}\\
0 &\text{otherwise}
\end{cases}
\end{align}
and for $i \in [k+1, m]$,
\begin{equation} \label{eq:Di_sol}
D_i^* = 1 - \sum_{j=k}^{i-1} f_j
\end{equation}
It could be observed that $D_{i+1}^* = a(i)$ which is the average number of nodes increasing degree from $i$ to $(i+1)$ when a node joins the network in the growth model \cite{scalefree:grow}. This is because, intuitively, the decreasing degree is exactly the opposite to a new node's connecting to $k$ neighbors.

The solution to Eq. [\ref{eq:vec}] may not be unique, an optimized solution with minimum message cost is discussed in Section \ref{sec:opt}.

\subsection{Protocol Design}
According to the analysis in Section \ref{sec:analysis}, there should be $I_i$ neighbors of $R$ that PUSH on degree $i$ and $D_i$ neighbors of $R$ that SHUFFLE on degree $i$, for $i=k,\ldots,m$. For the specific protocol design, we use the solution $I_i^*$ and $D_i^*$ presented in Eqs [\ref{eq:sol1},\ref{eq:Di_sol}].

%\textcolor{red}{[Deleted because it may distract readers]: Before node $R$ voluntarily quits, it can compute $I_i^*$ and $D_i^*$ locally and assign either the PUSH operation or the SHUFFLE operation to each of its neighbors. The neighbors will PUSH or SHUFFLE as assigned by $R$ before it quits.}

%\textcolor{red}{If node $R$ crashes due to errors or attacks, it can assign its neighbors the associated operations before the failure.}

It is necessary for every neighbor of R to know operations needed when R leaves the network, regardless if the departure is voluntary or caused by a crash. To ensure this, R should assign such operations to its neighbors as early as possible because it may accidentally crash anytime. As shown in Section \ref{sec:analysis}, the operations which will be assigned depend on the degree of R because the value of $I_i^*$ depends on the degree of $R$. Therefore, $R$ should re-compute $I_i^*$ and assign the new PUSH/SHUFFLE operations to the neighbors if its degree changes; in such a case the node sends \textit{updates messages} which contain its own degree and PUSH and SHUFFLE operations to its neighbors.

If the degree of a node $R$ changes to be $b'$, $R$ computes two non-decreasing sequences,
\begin{equation} \label{eq:uandv}
    v_i = \sum_{j=k+1}^{i} \frac{D_j^*}{k} \quad \textrm{and} \quad
    u_i = \sum_{j=k}^{i} \frac{I_j^*}{b'-k}
\end{equation}
and generates two random values $r_1, r_2$ distributed uniformly over the range $[0, 1)$ for each of the $b'$ neighbors. If $r_1 < k/b' $ and $r_2$ is in the interval $[v_i, v_{i+1})$, then the neighbor SHUFFLEs on degree $i$; if $r_1 \geq k/b' $ and $r_2$ is in the interval $[u_i, u_{i+1})$, then the neighbor PUSHes on degree $i$. In this way, the probability of PUSH on degree $i$ is $(1-\frac{k}{b'}) \frac{I_i^*}{b'-k} = \frac{I_i^*}{b'}$ and the probability of SHUFFLE on degree $i$ is $ \frac{k}{b'} \frac{D_i^*}{k} = \frac{D_i^*}{b'}$.

Furthermore, node $R$ sends the list of its neighbors' IPs in the \textit{update message}. When $R$ is removed from the network, all neighbors of $R$ broadcast the last \textit{update message} sent by $R$. Any receiver with one of the desired degrees in the \textit{update message} connects to the corresponding neighbor of $R$. And each neighbor of $R$ connects only to the first responder. It is worth noting that only one broadcast is needed because all neighbors of $R$ broadcast an identical message and nodes in the network forward the first message they receive.
 
The frequency of \textit{update message}s being sent significantly influences the efficiency and effectiveness of our protocol. If the messages were sent too frequently, the channels would be overwhelmed by \textit{update message}s, delaying regular traffic. Thus, our protocol should only use a limited number of \textit{update message}s. It is particularly important when nodes join and leave the network frequently. Here, we show that $M_{ave}$, the average number of the \textit{update message}s sent by a node during its lifetime, is bounded linearly by the maximum degree $m$ if nodes are randomly removed. Specifically, 
\begin{equation} \label{eq:updatedegree}
 M_{ave} \leq 3(m - 1 + k)k
\end{equation}
where the minimum degree $k$ is a small constant. The details of derivation of this bound are given in Appendix A.

Applying the solution from Eqs [\ref{eq:sol1},\ref{eq:Di_sol}] to our protocol, we obtain a new Enhanced Semi-Randomized Growth Algorithm: 
\\

\textit{Approach: E-SRA (Assuming one node is removed at a time)} To remove a node $R$ with degree $b$, each neighbor of $R$ either SHUFFLEs on degree $i$ by the probability $\frac{1 - \sum_{j=k}^{i-1} f_j}{b}$ for $i=k+1,\ldots, m $ or PUSHes on degree $j$ with probability $\frac{1}{b}$ for $j=k,\ldots, b-1$.\\

\begin{algorithm}
\caption{Enhanced Semi-Randomized Growth Algorithm (E-SRA)}\label{alg:1}
\begin{algorithmic}[1]
\Procedure{OnQuit}{}
\State Quit
\EndProcedure
\Procedure{OnDegreeChange}{newDegree}
\State $b \gets$ newDegree 
 \State \textit{update message} $\gets \{\}$
  \For { $i=1$ to $b$} 
  \State $r_1 \gets$ random number in $[0, 1)$
  \State $r_2 \gets$ random number in $[0, 1)$
  \If {$r_1 < \frac{k}{b}$}
    \For { $j=k+1$ to $m$ }
      \If {$r_2 \in [v_j, v_{j+1})$}
        \State $op_i \gets$ ( ${\text{IP}}_i$ , \textit{SHUFFLE on degree j})
      \EndIf
    \EndFor
  \Else
    \For { $j=k$ to $m-1$ }
      \If {$r_2 \in [u_j, u_{j+1})$}
        \State $op_i \gets$ ( ${\text{IP}}_i$ , \textit{PUSH on degree j})
      \EndIf
    \EndFor
  \EndIf
  \EndFor
  \State \textit{update message} $\gets \{op_1,op_2\ldots,op_b,b\}$
  \State Send the \textit{update message} to every neighbor
\EndProcedure
\Procedure{OnNeighborQuit}{}
\State Broadcast the latest \textit{update message} received from the quitting neighbor
 \State Connect to the first responder
\EndProcedure
\Procedure{OnReceiveUpdateMessage}{updateMsg}
\State $b \gets$ the degree of itself
 \For { each $op_i$ in updateMsg}
  \If { $op_i =$ ( ${\text{IP}}_i$ , \textit{PUSH on degree $b$}) }
    \State Reply to the node with ${\text{IP}}_i$
    \State Return
  \ElsIf { $op_i =$ ( ${\text{IP}}_i$ , \textit{SHUFFLE on degree $b$}) }
    \State Terminate its existing connection to a random neighbor RN
    \State Ask RN to reply to the node with ${\text{IP}}_i$ 
    \State Return
  \EndIf
 \EndFor
\EndProcedure

\end{algorithmic}
\end{algorithm}

The pseudo-code is listed as Algorithm \ref{alg:1}. The message complexity of the algorithm is one broadcast per removal.

As illustrated in Algorithm \ref{alg:1}, if a node's degree is changed, it sends an \textit{update message} to its neighbors (as shown in lines 3-18) so that they know to which nodes to rewire if the failure occurs. The operations in the \textit{update message} are computed by generating a sequence of random numbers (as shown in lines 7-16). Since a node knows its operations in advance, if a neighbor fails, this node will operate as what it was assigned (as shown in lines 19-21). And the design of this protocol does not require any extra steps for a node gracefully quitting (as shown in lines 1-2) because the neighbors of this quitting node already know the desired degrees of the nodes to which they should rewire afterwards.

In Eq. [\ref{eq:Di_sol}], the greater $i$ is, the smaller $D_i^*$ is. This fact makes the protocol more practical because there are a large number of nodes with low degrees and only a small fraction of nodes with high degrees in a scale-free network. Nodes with low degrees are easily reached by broadcasts in a few hops; Eq. [\ref{eq:sol1}] implies that the number of PUSH operations increases linearly with the degree of node $R$.

Note that we assume the neighbors of node $R$ are able to PUSH or SHUFFLE when $R$ is removed. When two connected nodes voluntarily quit at the same time, one node should wait until the other node quits successfully (the tie can be broken by the order of IP addresses); if a node crashes while its neighbors are alive, the neighbors can also PUSH or SHUFFLE correctly. However, the above algorithm does not work if node $R$ and its neighbor crash at the same time because both crashed nodes can neither PUSH nor SHUFFLE. In this paper, we assume that the nodes crash incrementally, one after the other; in the case that a group of connected nodes crash simultaneously, a global restoration algorithm is more appropriate to use.

\subsection{Optimization of the Message Costs} 
\label{sec:opt}
The number of replies to the neighbors of $R$ can be optimized by applying better solutions satisfying Eq. [\ref{eq:vec}]. 

Let $s_i$ denote the number of nodes with degree $i$ for $i\in[k,m]$ among the neighbors of node $R$ (thus, $\sum_{i=k}^{m} s_i = b$). If a neighbor of $R$ is assigned to SHUFFLE on degree $i$ and this neighbor has $i$ connections, then it does not broadcast because it lost the connection to $R$, so its degree decreases to $(i-1)$, matching the desired decrease of degree. Taking this into consideration, one SHUFFLE on degree $i$ can be skipped if $R$ has a neighbor with degree $i$. The average number of the skipped SHUFFLEs is $\sum_{i=k+1}^{m} |{s_i - D_i}|$. Using Eq. [\ref{eq:vec}] as the constraint, the optimization problem can be formulated over $\vec{I}$ and $\vec{D}$ as,
\begin{align}
& \underset{\vec{I} \geq 0, \vec{D} \geq 0}{\text{min}}
& &   \| \vec{S} - \vec{D} \|_1 \\
& \text{subject to}
& & \vec{I} - \vec{D} =  \vec{C_b}\\
&&& \| \vec{I} \|_1 = b-k \\
&&& \| \vec{D} \|_1 = k 
\end{align}
where $\vec{C_b} = (f_k, f_k+f_{k+1}, \ldots, \sum_{i=k}^{b} f_i - 1, \ldots, \sum_{i=k}^{m-1} f_i - 1)^T$ and $\vec{S} = (s_{k+1}, s_{k+2},\ldots, s_{m})$. This problem can be rewritten as linear programming problem and be solved by the linear programming technique efficiently in real time.

The optimal solution depends on the degree distribution of $R$'s neighbors. So, node $R$ computes the new optimal solution and assigns new atomic operations to its neighbors via sending an \textit{update message} if its degree changes.

\begin{table}[!t]
\caption{Comparison of Growth Models}
\label{table:compare}
\centering
\setlength\tabcolsep{2pt}
\begin{tabular}{|c|c|c|c|}
\hline
Algorithm&Global knowledge used&Flexible $\gamma$& Tolerance to removal\\
\hline
\hline
BA\cite{scalefree:BA} & Complete & No & No\\
\hline
HAPA\cite{scalefree:yuksel} & Partial & No & No\\
\hline
Gaian\cite{scalefree:Gaian} & None & No & No\\
\hline
subPA\cite{yuksel:adhoc} & $\tau_j$-hop and $\tau_l$-hop neighbors & No & Yes\\
\hline
SRA\cite{scalefree:grow} & None & Yes & No\\
\hline
E-SRA & None & Yes & Yes\\
\hline
\end{tabular}
\end{table}

\section{Simulations} \label{sec:sim}
Empirical experiments are conducted to evaluate the overlay topologies produced by E-SRA in various settings. These topologies are compared with those produced by HAPA, Gaian and subPA algorithms using the same configurations. We measure the fitness of the produced degree distribution to the power-law and then evaluate the search efficiency over these topologies by running Flooding (FL) algorithm and Normalized Flooding (NF) algorithm, which are simple search algorithms used in unstructured P2P networks\cite{search:1}.

\subsection{Simulation Settings}
By adding and removing nodes dynamically, a set of network topologies are produced with different parameters. At the beginning, a network of $(2k+1)$ nodes is constructed. Each node connects to all the other $2k$ nodes so that the average degree is $2k$. For all simulations, 5,000 nodes are added first and then we run 145,000 iterations of joining and removing nodes to produce one topology. In each iteration, either a new node joins the network or an existing node is removed. In order to simulate the dynamics of nodes joining and leaving in real-world applications, a node joins with probability $(1-p)$ and quits with probability $p$ in one iteration. Thus, a few nodes are quite likely to join or leave the network in a sequence of iterations. The value of $p$ is smaller than $1/2$ to keep the network growing. In a topology produced by E-SRA, nodes with desired degrees may not exist when the network size is small. In such cases, the nodes can rewire to the existing nodes with any degree\footnote{In our simulations, this case occurred in only 3\% of the deletions until the first network topology obtained with all degrees.}. However, with the growth of the network, there will be sufficient nodes with all possible degrees.

Experiments have been conducted on the topologies produced by HAPA, Gaian and subPA algorithms with the same pattern of nodes joining and leaving the network. After a sufficient number of such operations, the degree distribution is calculated based on a "snapshot" of the network topology and is compared with the perfect power-law distribution. Since all four algorithms utilize randomized approaches, we take the average of the degree distribution of 10 randomly produced topologies with the same parameters to study the average cases. 

Search algorithms are implemented to test the search efficiency over the produced topologies in different settings. We consider two search algorithms in P2P networks: 1) Flooding (FL), where every node forwards a query to all neighbors until the query hits the target. 2) Normalized Flooding (NF), where every forwarder randomly chooses $k$ (i.e. the minimum degree) neighbors and sends them the query. Time to live (TTL), which is the maximum number of hops a message can traverse, is set up to limit the lifetime of a query in a network. So, a query either reaches its destination or expires due to its TTL. It is assumed that the message sources are uniformly distributed in the network.

Gaian\cite{scalefree:Gaian} and HAPA\cite{scalefree:yuksel} algorithms do not specify how nodes are removed so their nodes with the minimum degree $k$ are likely to lose connections. This results in appearance of nodes with the degree smaller than the minimum degree $k$. To avoid such effect, we assume the node with degree smaller than $k$ will connect to existing nodes using the original approach to regain $k$ connections. Nodes with the maximum degree $m$ will not accept new connections. The hard cut-off is implemented as in the original approaches and will not be influenced by the effect of removing nodes. In subPA, the parameter used to constructed subset of nodes is $\tau_j = 2$, $ \tau_l = 2$.

\begin{figure*}[!t]
\centering
\subfloat[E-SRA  $\gamma=2.5$]{\includegraphics[width=2.3in]{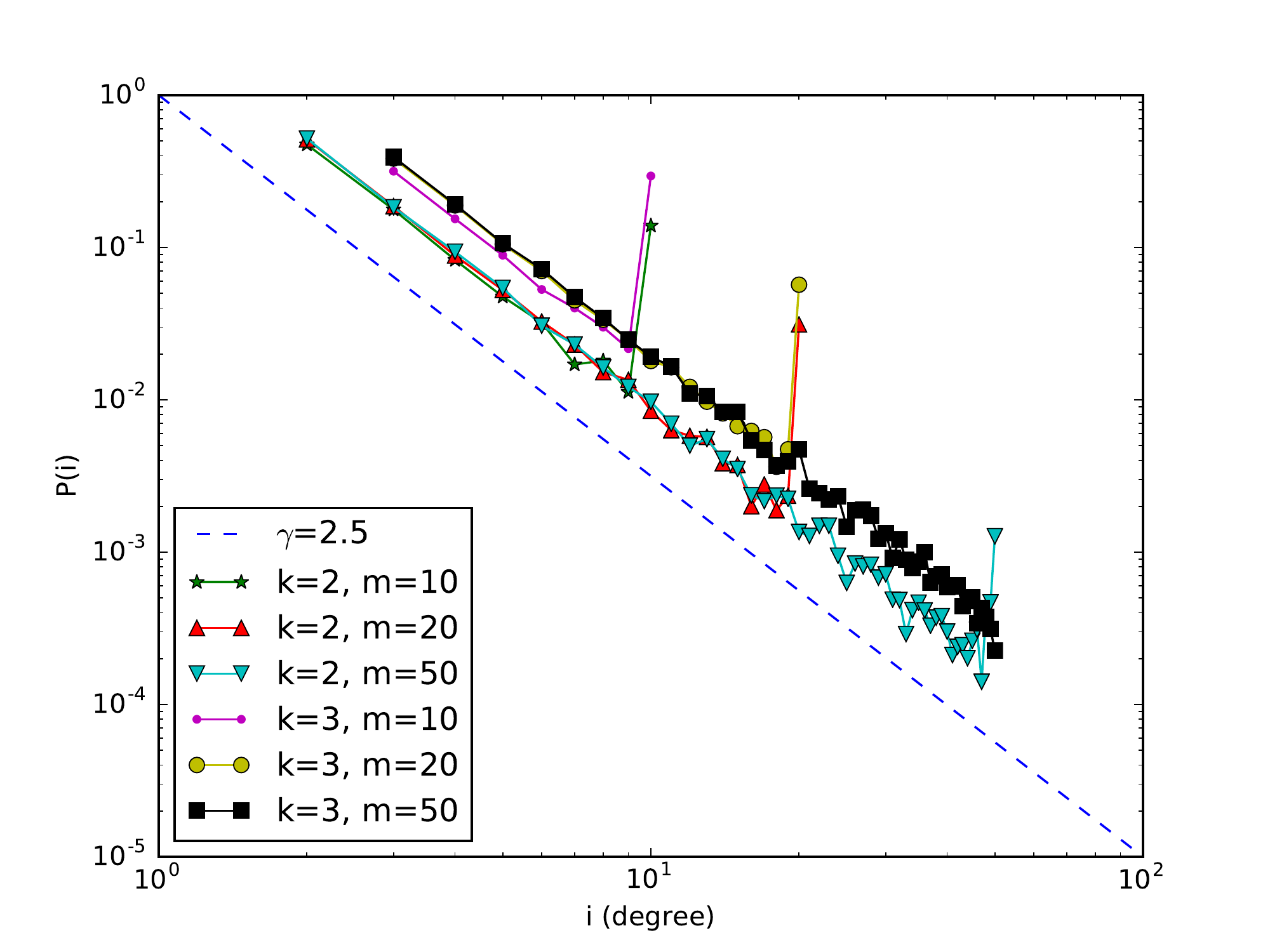}%
\label{remove25}}
\hfil
\subfloat[E-SRA $ \gamma=2.7$]{\includegraphics[width=2.3in]{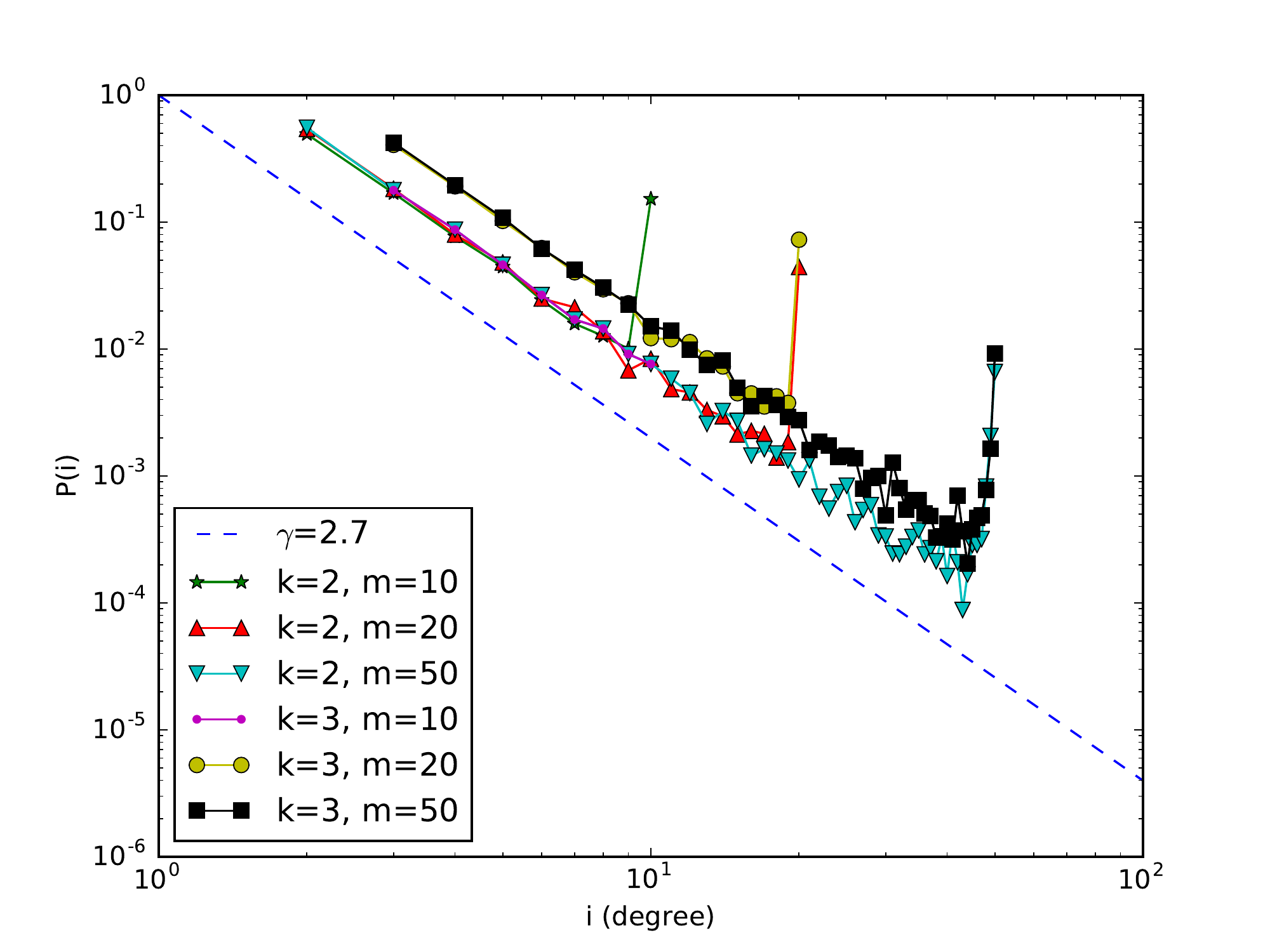}%
\label{remove27}}
\hfil
\subfloat[E-SRA $ \gamma=3$]{\includegraphics[width=2.3in]{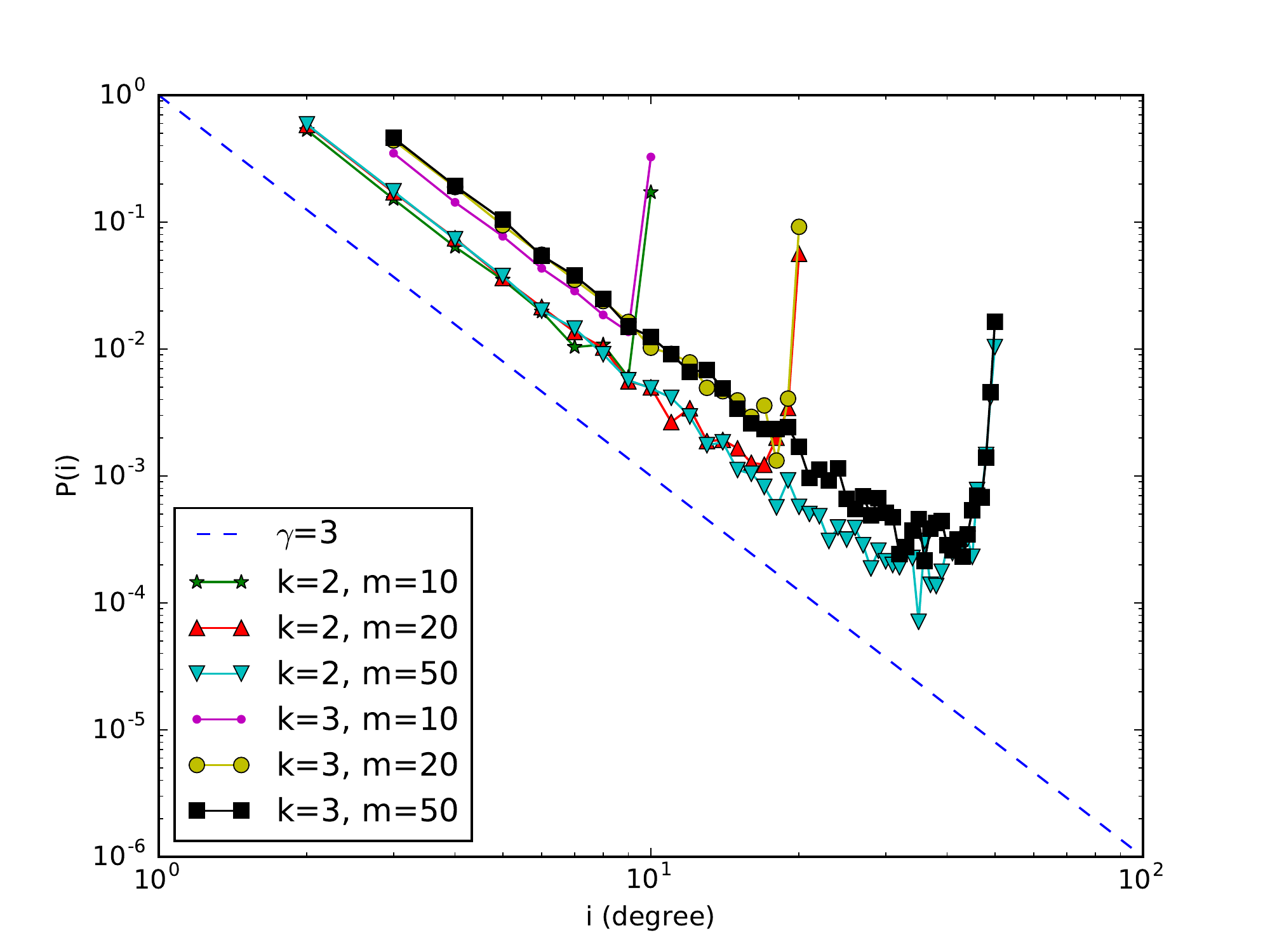}%
\label{fig:remove30}}
\hfil
\subfloat[Gaian]{\includegraphics[width=2.3in]{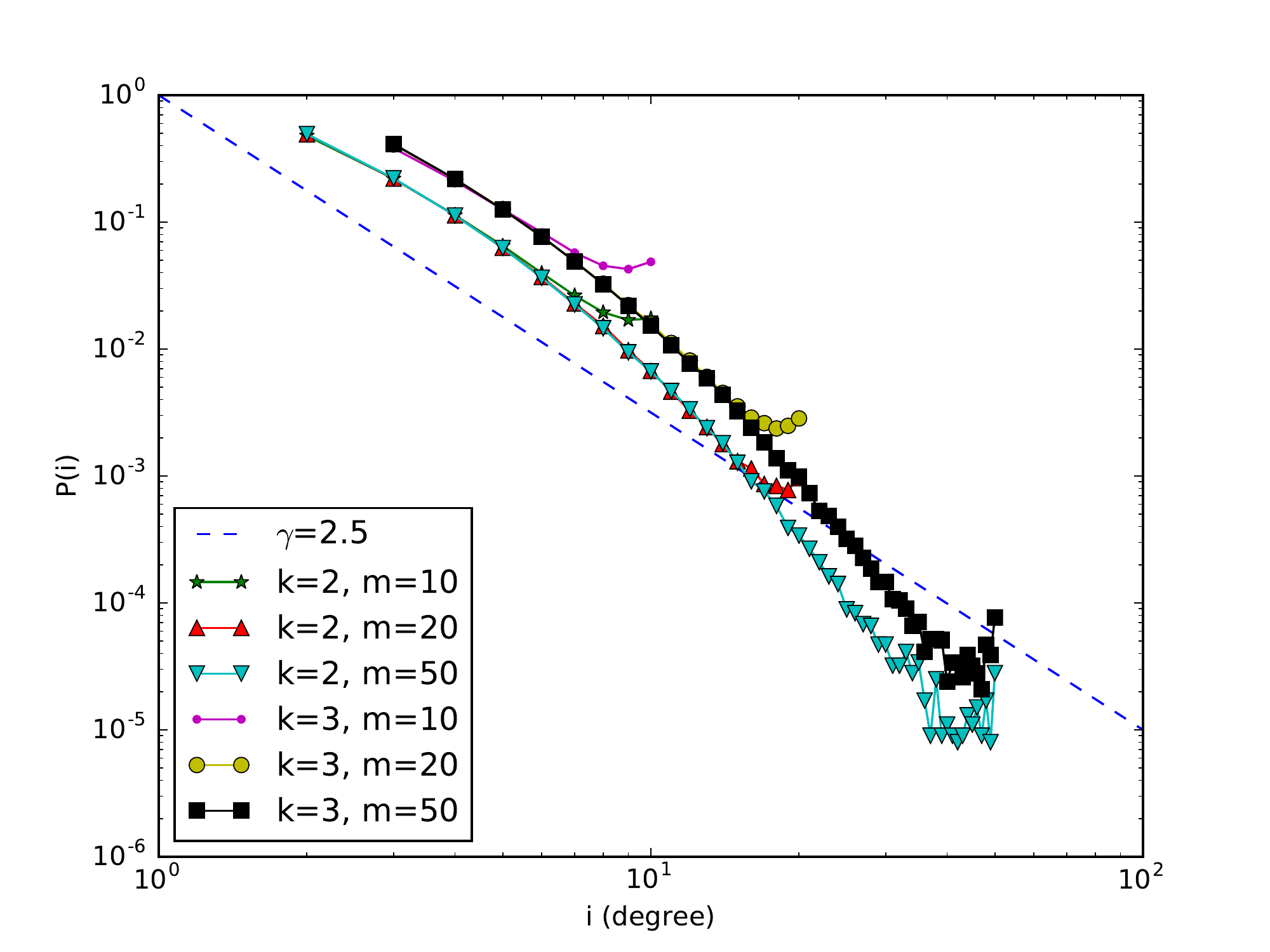}%
\label{gaian}}
\hfil
\subfloat[HAPA]{\includegraphics[width=2.3in]{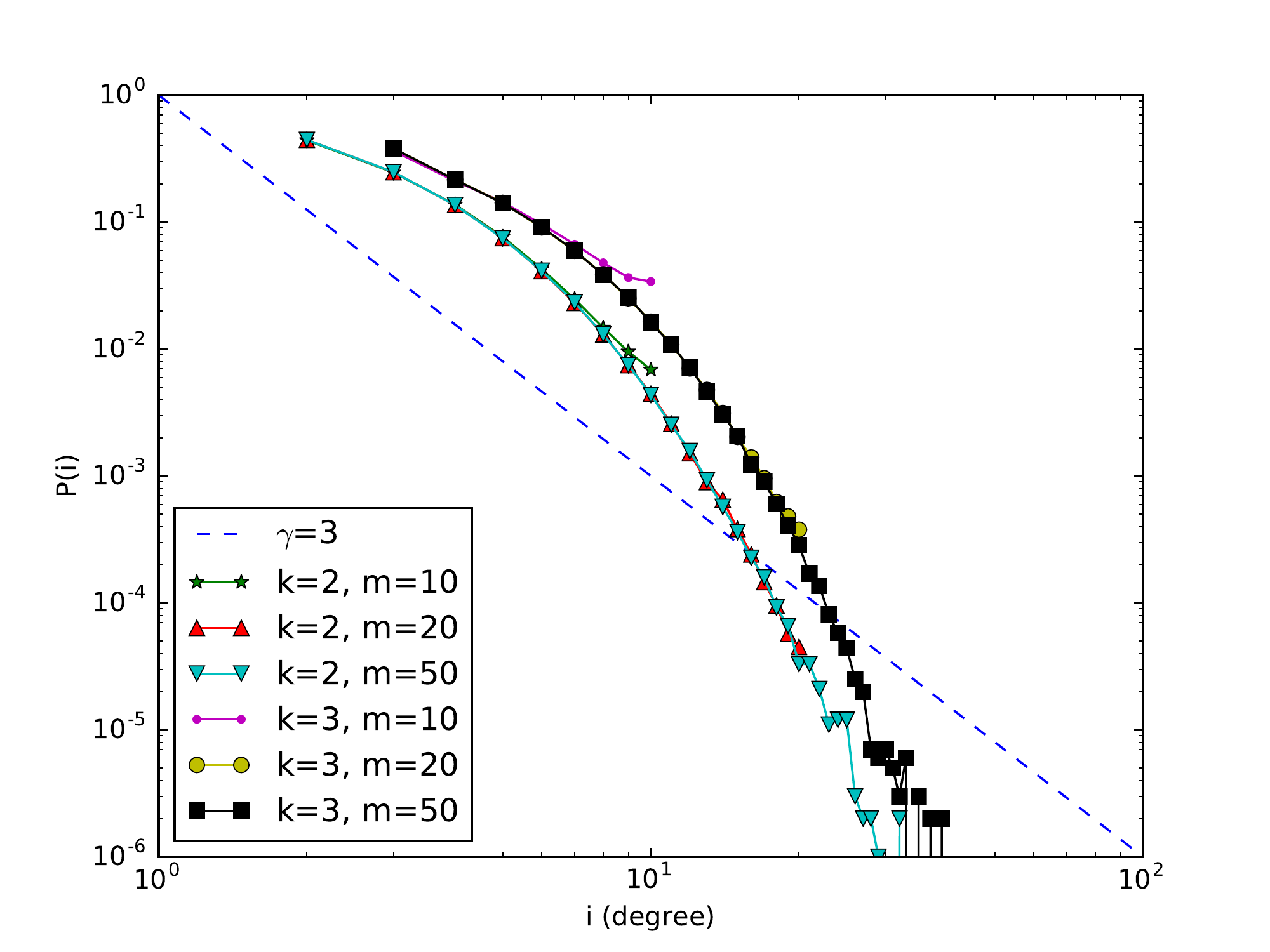}%
\label{hapa}}
\hfil
\subfloat[subPA]{\includegraphics[width=2.3in]{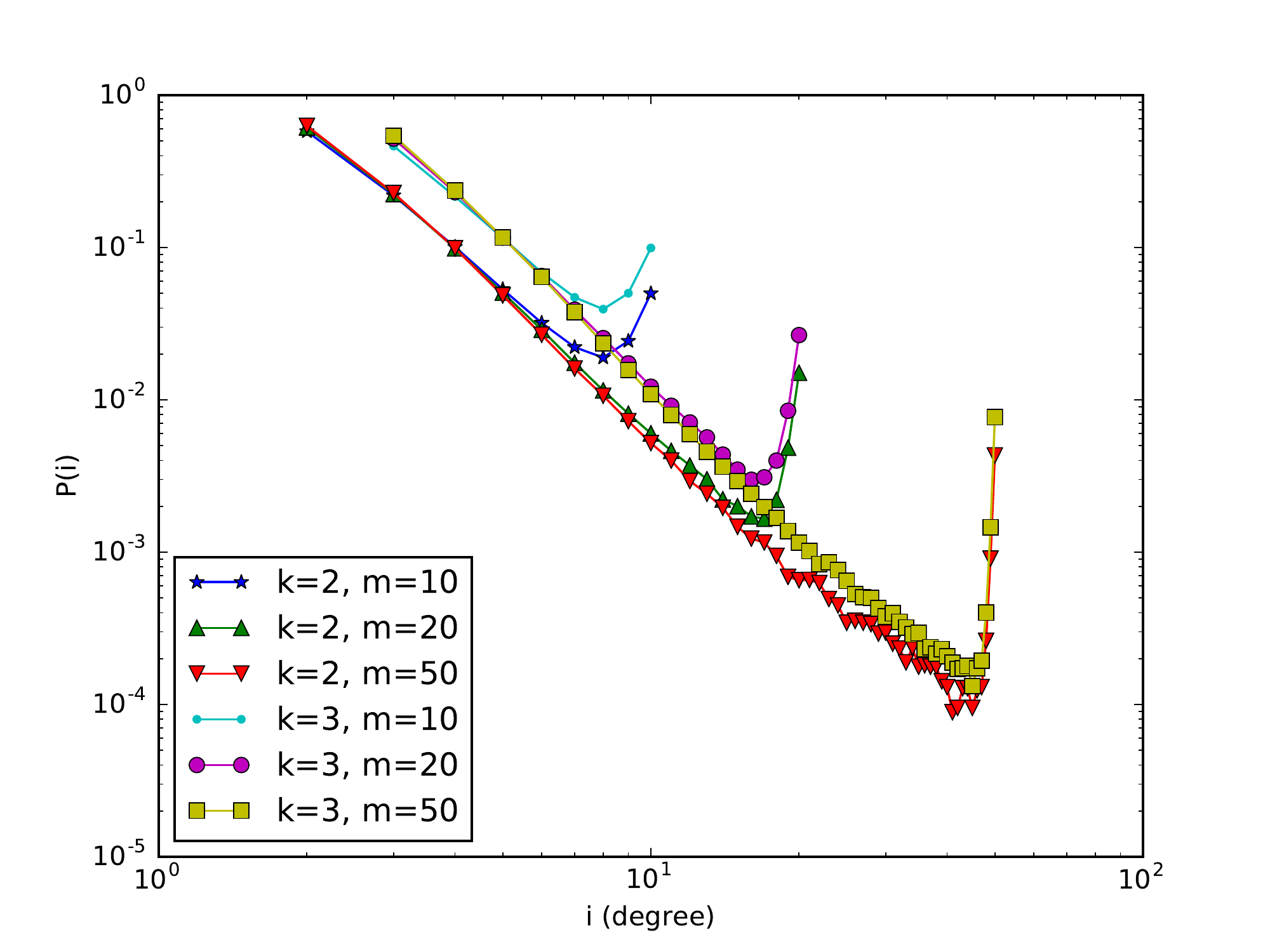}%
\label{subpa_random}}
\caption{Degree distribution of the topologies where nodes are randomly removed. $n \approx 50000$, $p=1/3$.}
\label{fig:fig_sim_random}
\end{figure*}

\subsection{Degree Distribution}
It can be observed that regardless of parameter settings E-SRA has produced topologies with the degree distribution perfectly matching the power-law. 

Figure \ref{fig:fig_sim_random} illustrates the degree distribution of the produced topologies when nodes are removed randomly. In each iteration, either an existing node quits with probability $p = 1/3$ or a new node joins the network with probability $1-p = 2/3$. The node to be removed is randomly chosen from the network. We apply a total of $1.5 \times 10^5$ iterations to produce the final topology. Thus every topology has approximately $(1-2p) \times (1.5 \times 10^5) = 5 \times 10^4$ nodes. In E-SRA, the scaling parameter is set as $2.5$, $2.7$ and $3$ respectively. It can be observed from Figure \ref{fig:fig_sim_random} that both Gaian and HAPA algorithms have produced a sufficient number of nodes with low degrees whereas the number of nodes with high degrees is insufficient. Since nodes with high degrees are more likely to connect to the nodes to be removed, their degrees decrease with high probability compared to nodes with low degrees. Compared to HAPA and Gaian algorithms, E-SRA and subPA have generated sufficient number of nodes with high degrees.

As the simulation results suggest, the power-law is approximately preserved in all four growth models compared here when the maximum degree is $m=10$. But, if the maximum degree is $m=50$, HAPA and Gaian algorithms produce fewer nodes with high degree than needed. This is because the number of nodes with high degrees are of the order of several thousand with $m=10$, but there are fewer than $100$ when $m=50$. So the final degree distribution is more sensitive to algorithm imprecision and the difference is easier to observe for m=50 than for m=10. For the same reason, E-SRA produces a topology with a small tail at degree $48$, $49$ when $m=50$. As the network size grows, the total numbers of nodes at all degrees increase and the tail disappears.

\begin{figure*}[!t]
\centering
\subfloat[E-SRA $\gamma=2.5$]{\includegraphics[width=2.3in]{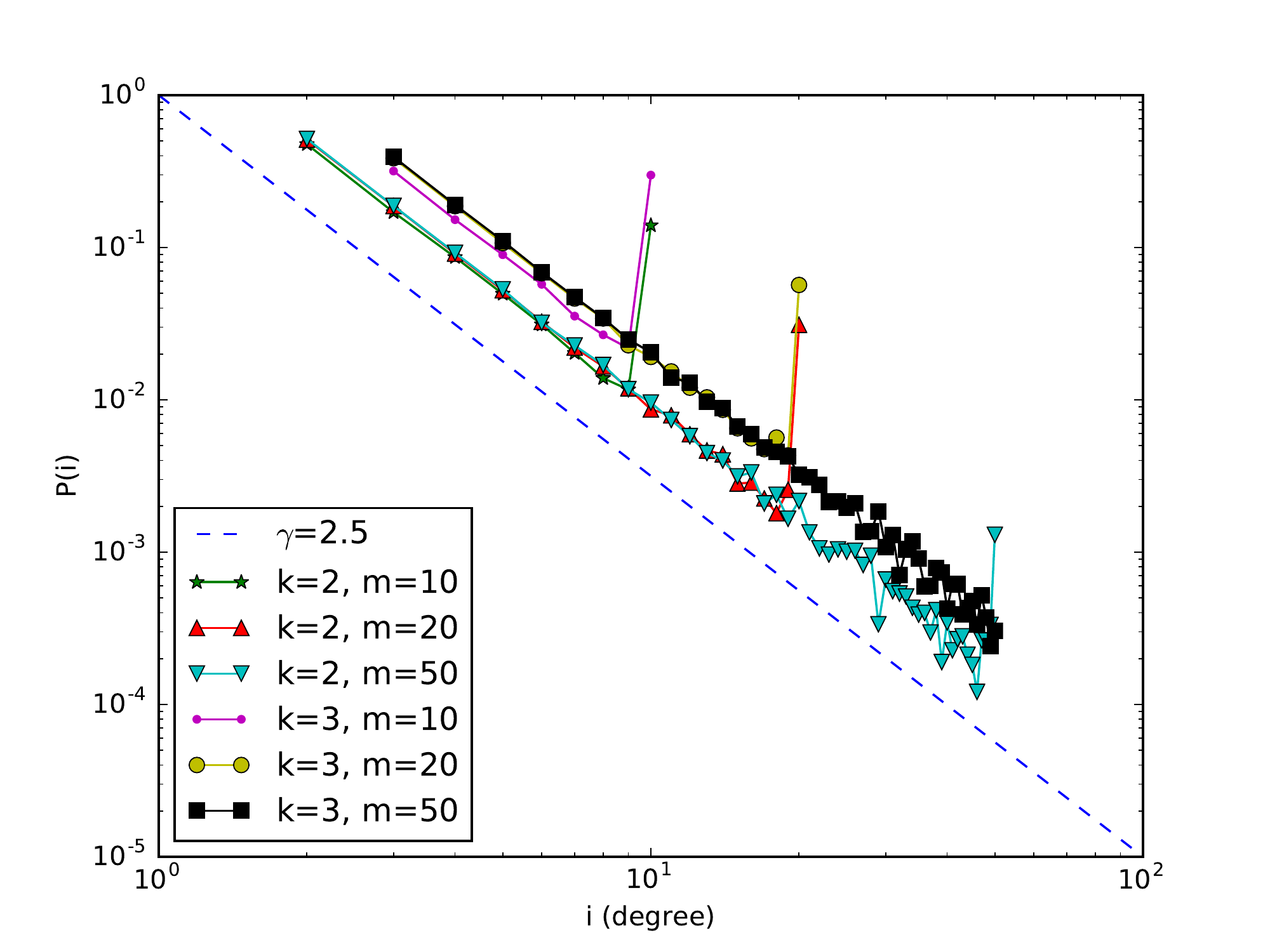}%
\label{remove25_hub}}
\hfil
\subfloat[E-SRA $\gamma=2.7$]{\includegraphics[width=2.3in]{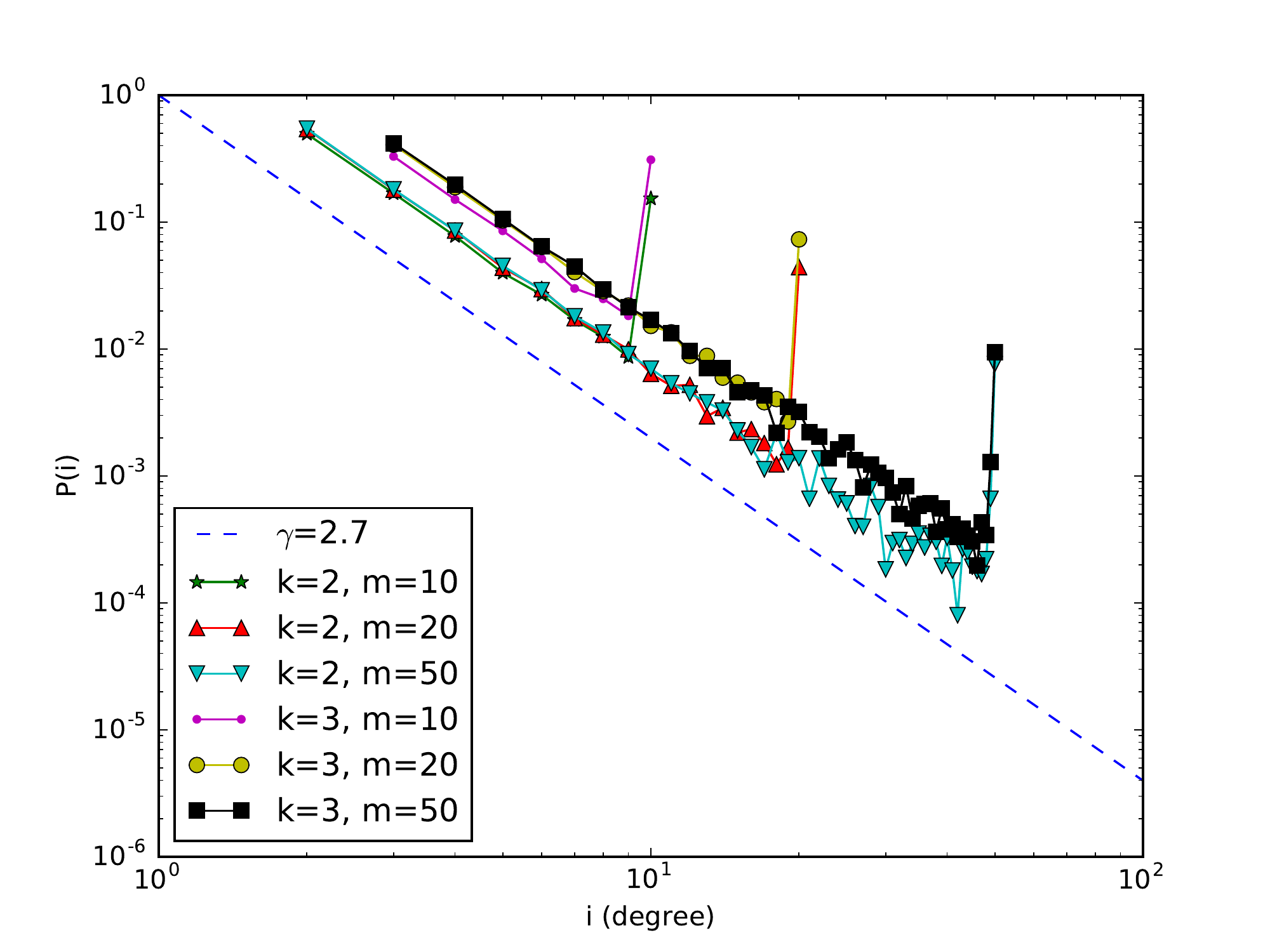}%
\label{remove27_hub}}
\hfil
\subfloat[E-SRA $\gamma=3$]{\includegraphics[width=2.3in]{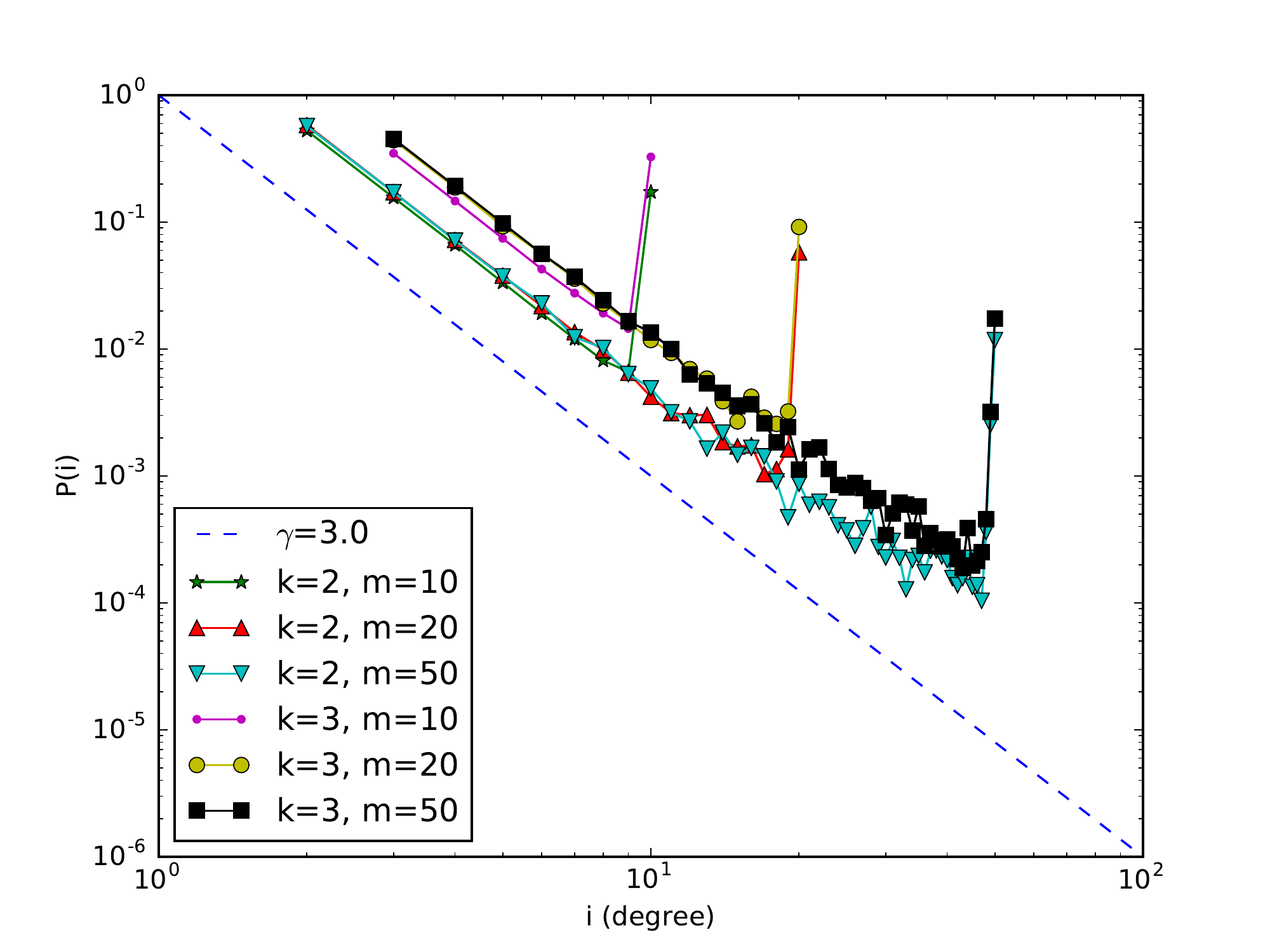}%
\label{remove30_hub}}
\hfil
\subfloat[Gaian]{\includegraphics[width=2.3in]{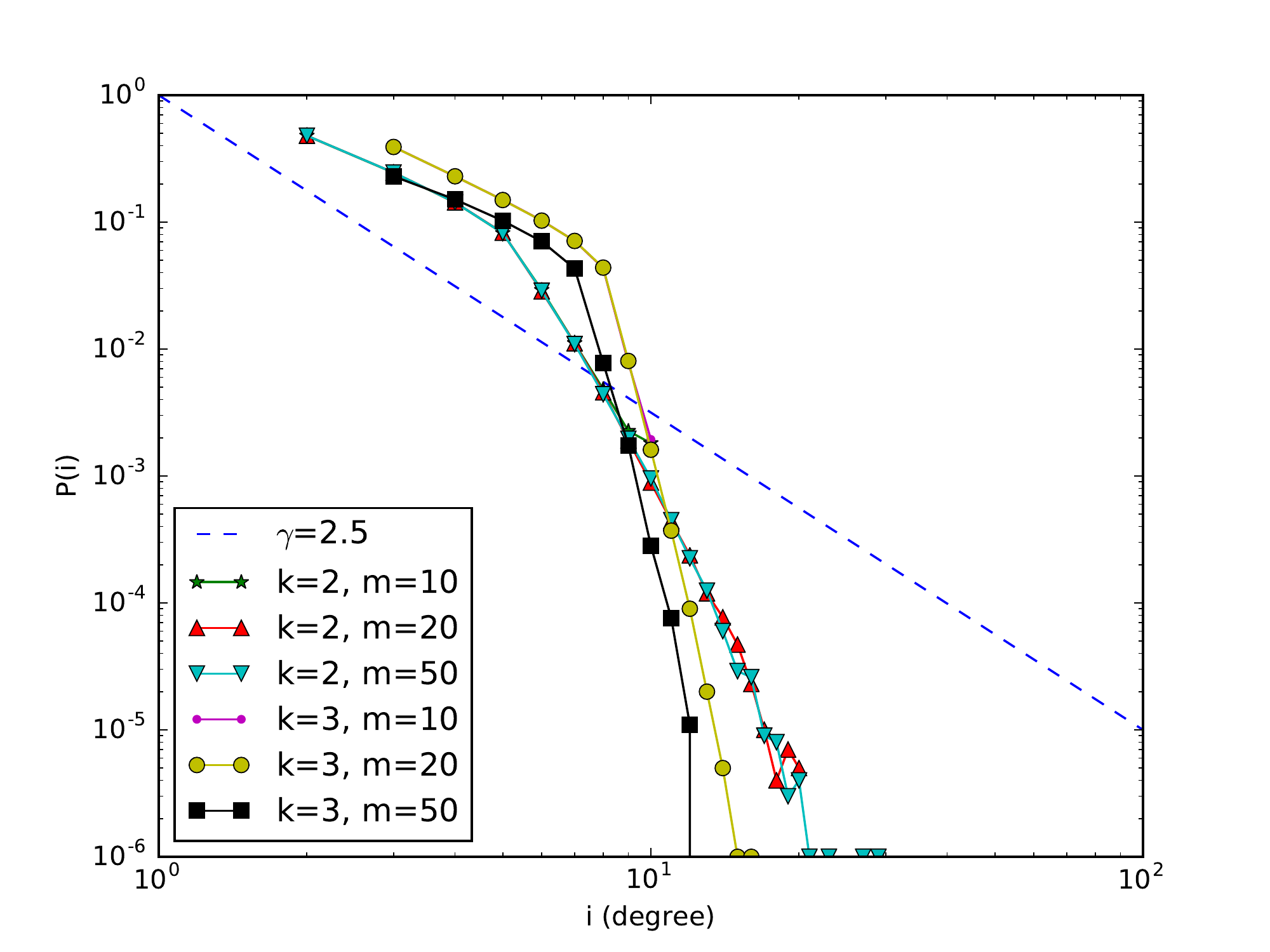}%
\label{fig:gaian_hub}}
\hfil
\subfloat[HAPA]{\includegraphics[width=2.3in]{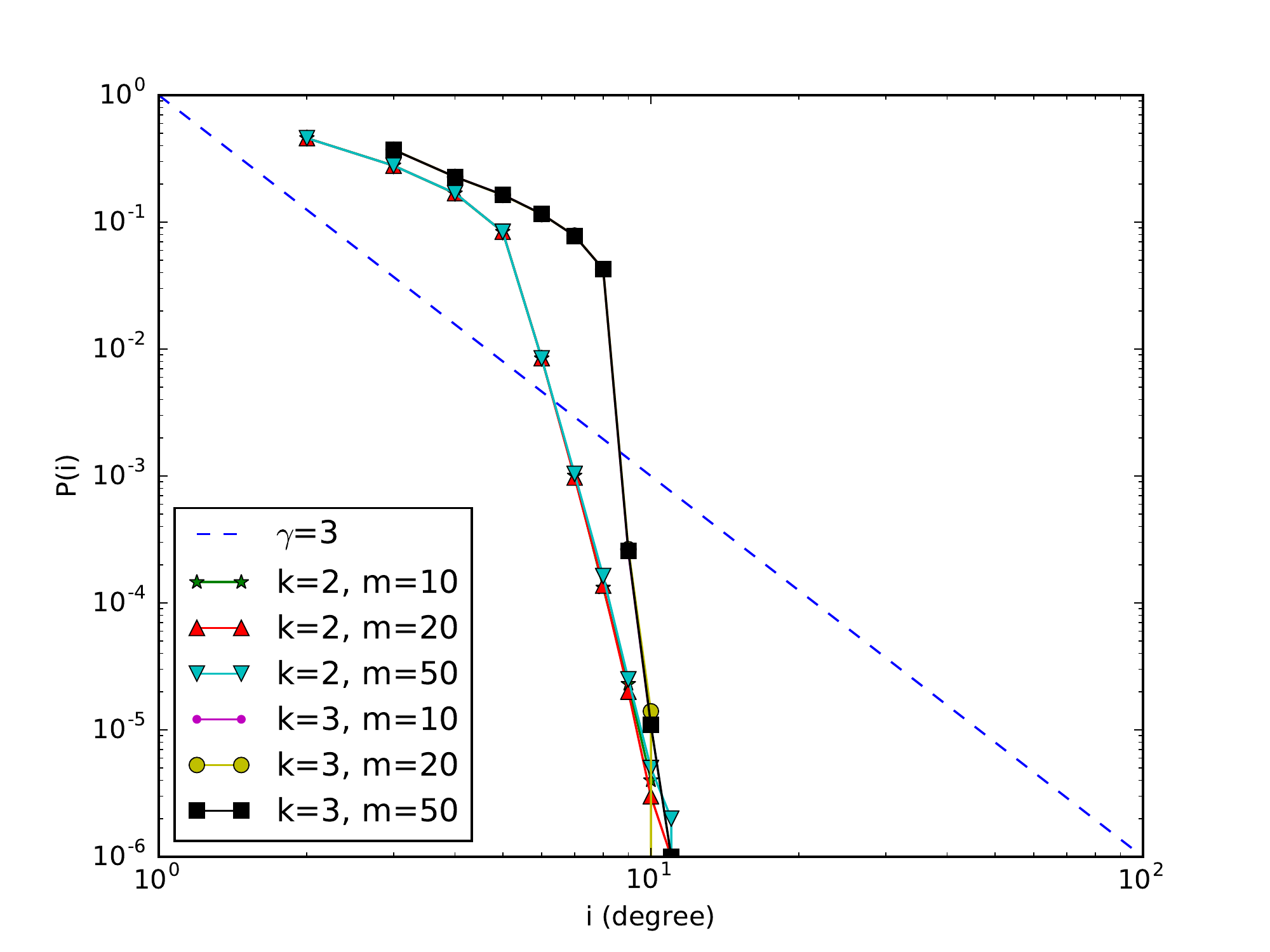}%
\label{fig:hapa_hub}}
\hfil
\subfloat[subPA]{\includegraphics[width=2.3in]{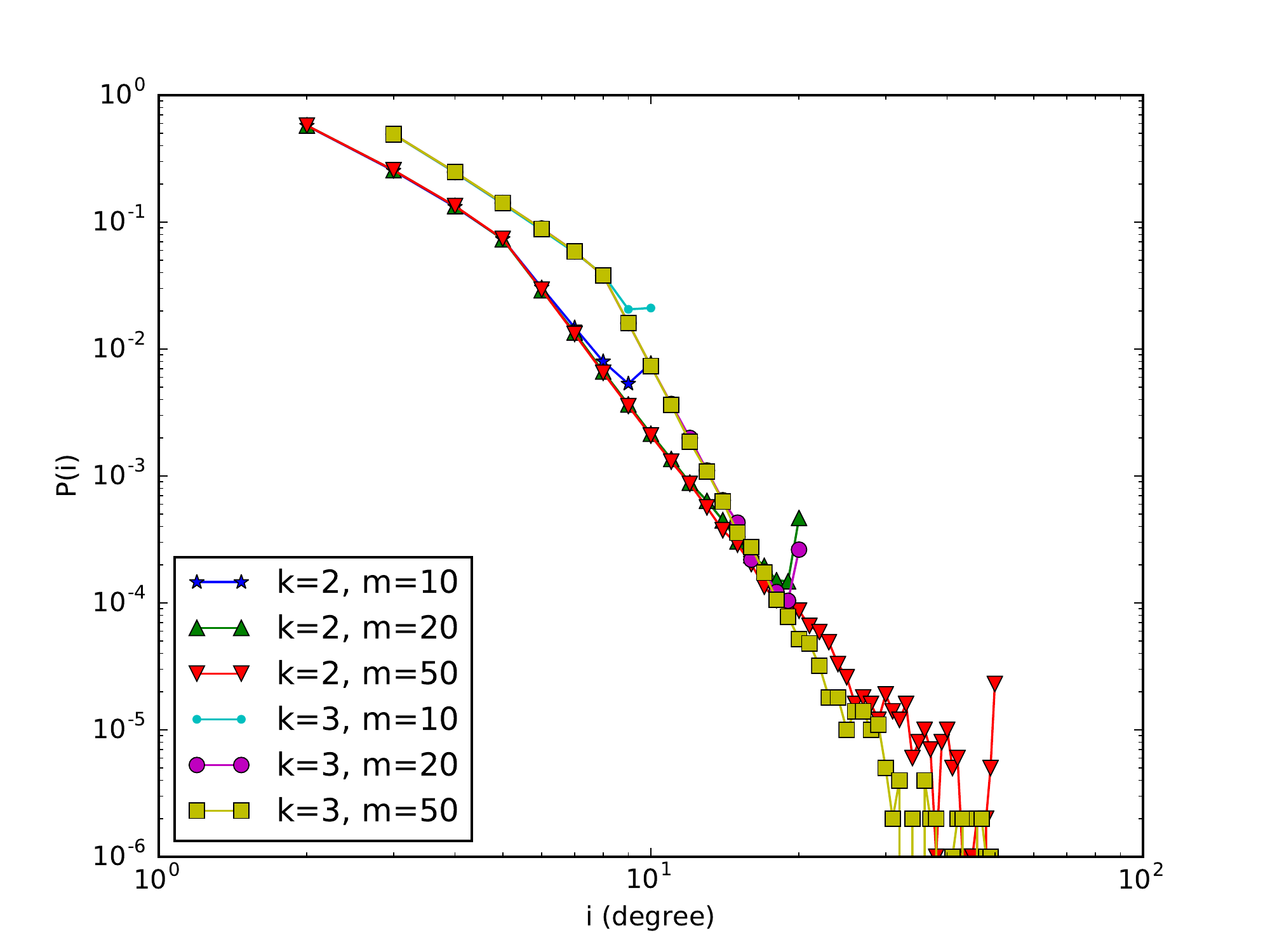}%
\label{fig:subpa_hub}}

\caption{Degree distribution of the topologies where nodes with degree at least $3k$ are randomly removed. $n \approx 90000$, $p=1/5$.}
\label{fig:fig_sim_hub}
\end{figure*}

Figure \ref{fig:fig_sim_hub} shows the simulation results when the nodes with degree at least $3k$ are randomly removed. Here, every iteration removes a node with degree at least $3k$ with probability $p = 1/5$ or adds a node with probability $1-p = 4/5$. This removal pattern simulates a topology with hubs leaving frequently. As shown in Figures \ref{fig:fig_sim_hub}.d and \ref{fig:fig_sim_hub}.e, the scale-free topologies constructed without a special mechanism handling the removal are very vulnerable to attacks on hubs. The topologies produced by Gaian and HAPA algorithms are significantly below the required number of nodes with high degrees when the maximum degree $m=20$ or higher. In HAPA, nodes with degrees close to the cut-off value completely vanish. The topologies with the maximum degree $10$ are a little more robust in subPA, HAPA and Gaian, but even in this case nodes with degrees of $9$ and $10$ are less numerous than what the power-law requires. In contrast to subPA, HAPA and Gaian algorithms, E-SRA manages to maintain the limited scale-free topology when nodes with degrees at least $3k$ are randomly removed.

\iffalse
Smaller probability of removal compared with previous experiment is chosen to \textbf{prevent exhausting****} nodes with degree at least $3k$.
\fi

\begin{figure*}[!t]
\centering
\subfloat[E-SRA $\gamma=2.5$]{\includegraphics[width=2.3in]{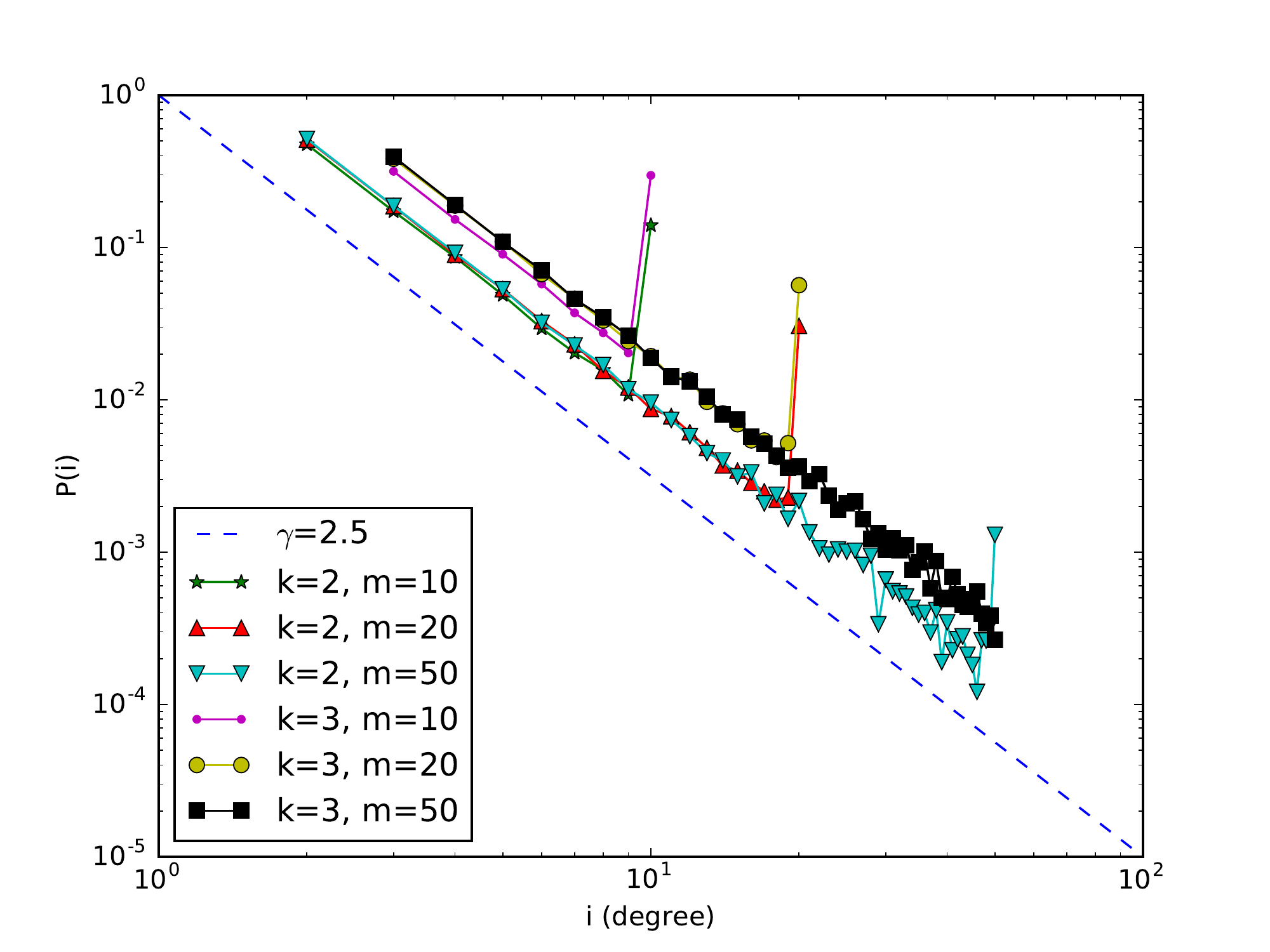}%
\label{remove25_less}}
\hfil
\subfloat[E-SRA $\gamma=2.7$]{\includegraphics[width=2.3in]{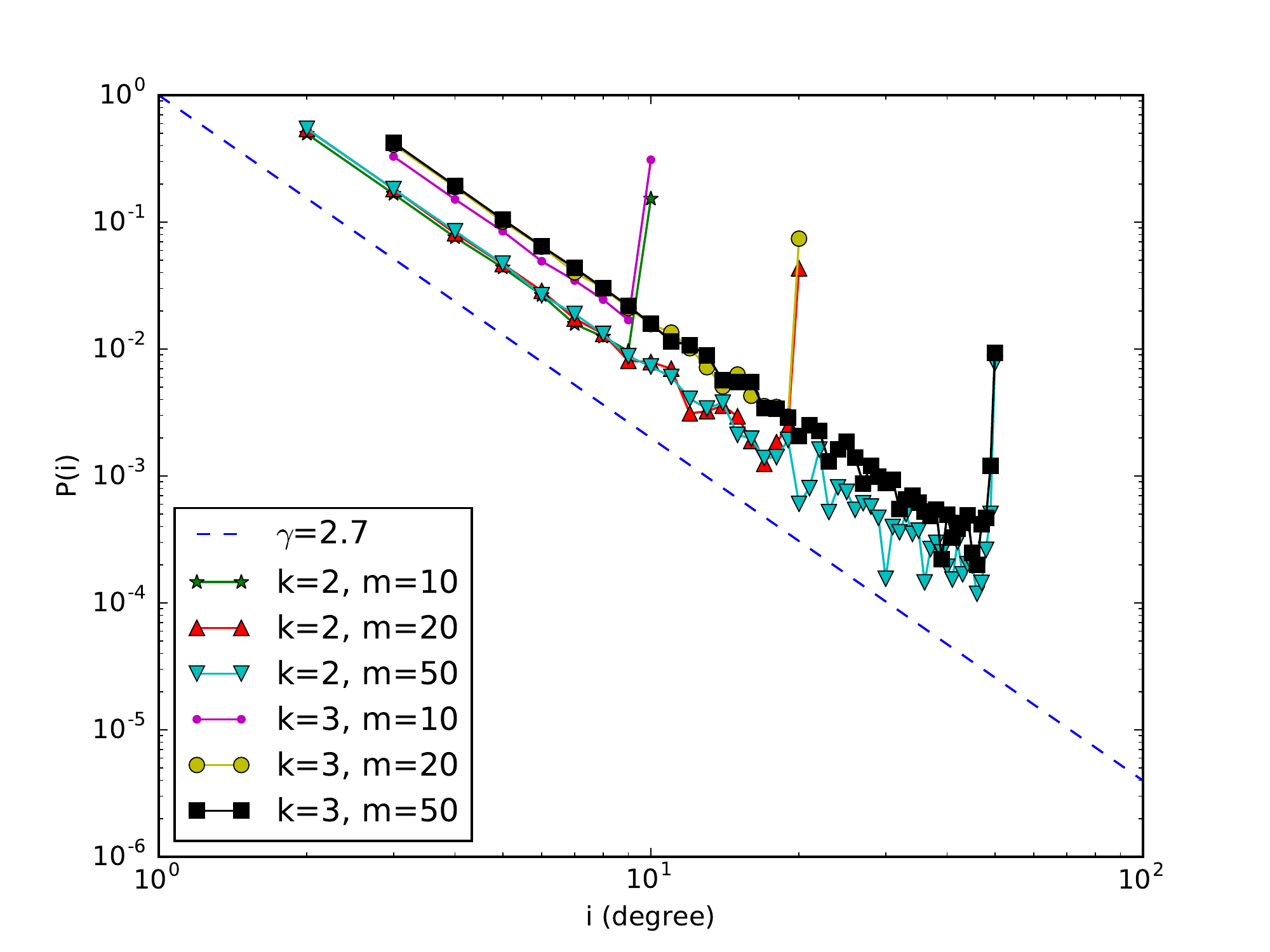}%
\label{remove27_less}}
\hfil
\subfloat[E-SRA $\gamma=3$]{\includegraphics[width=2.3in]{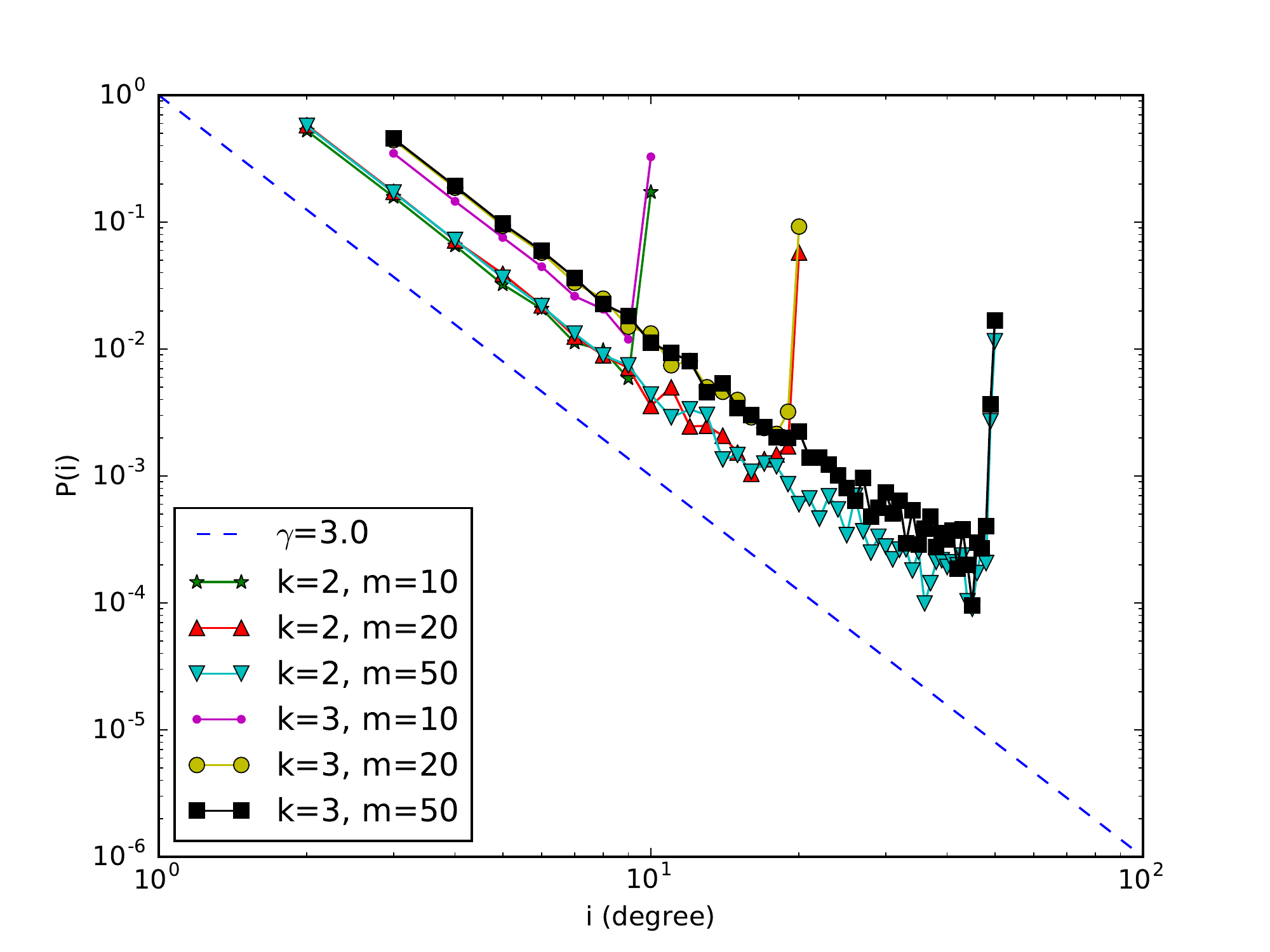}%
\label{remove30_less}}
\hfil
\subfloat[Gaian]{\includegraphics[width=2.3in]{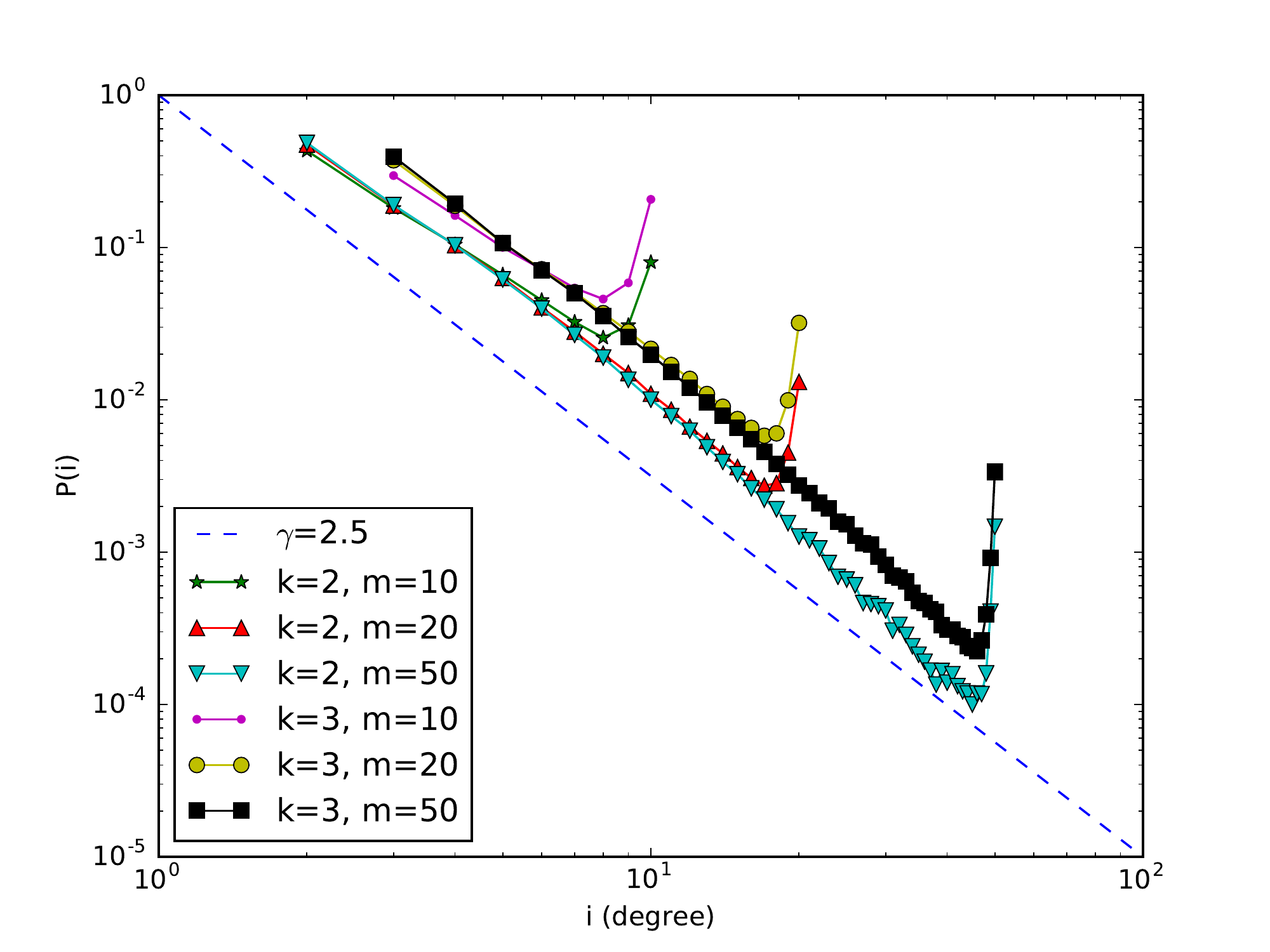}%
\label{gaian_less}}
\hfil
\subfloat[HAPA]{\includegraphics[width=2.3in]{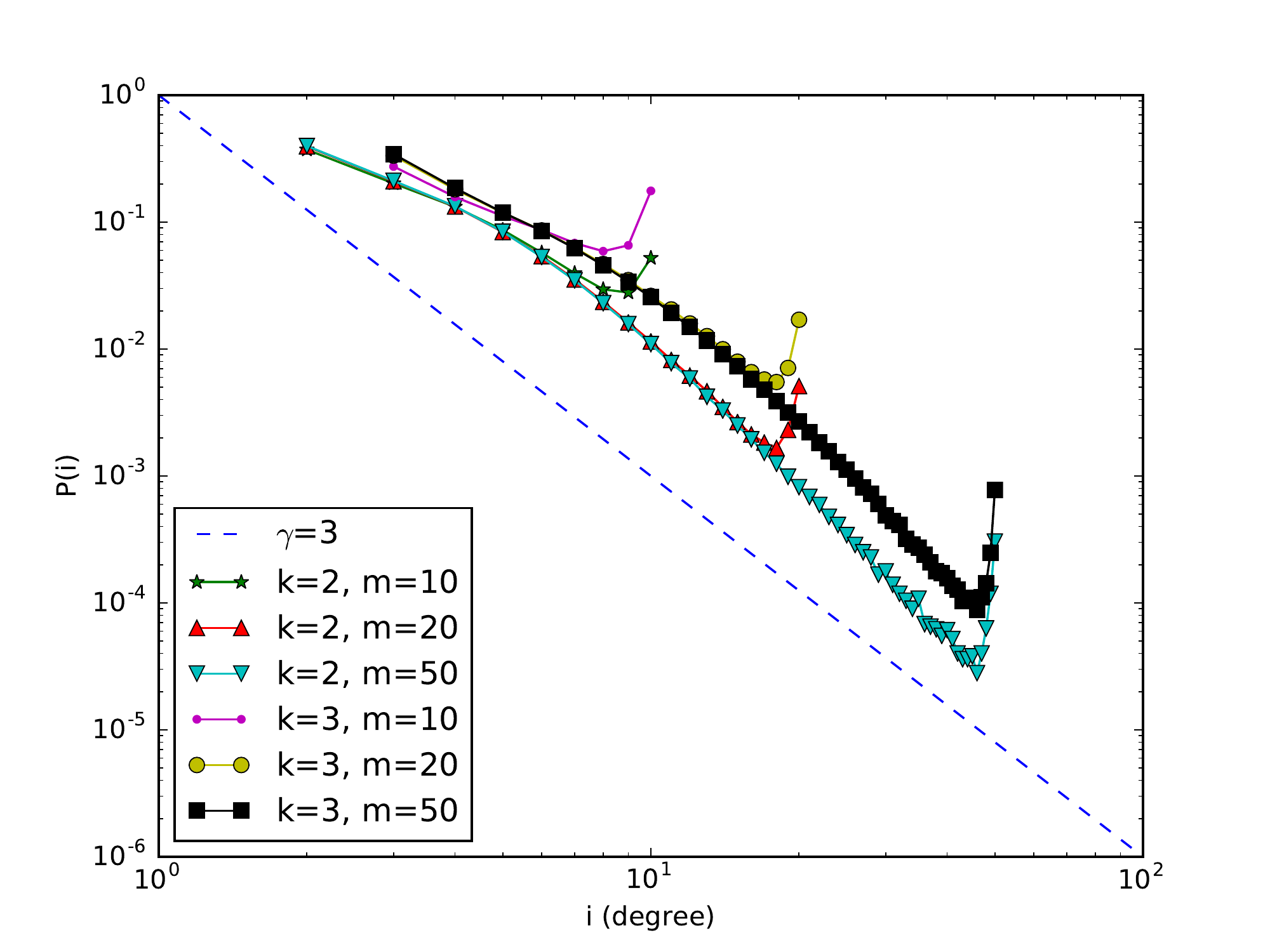}%
\label{hapa_less}}
\hfil
\subfloat[subPA]{\includegraphics[width=2.3in]{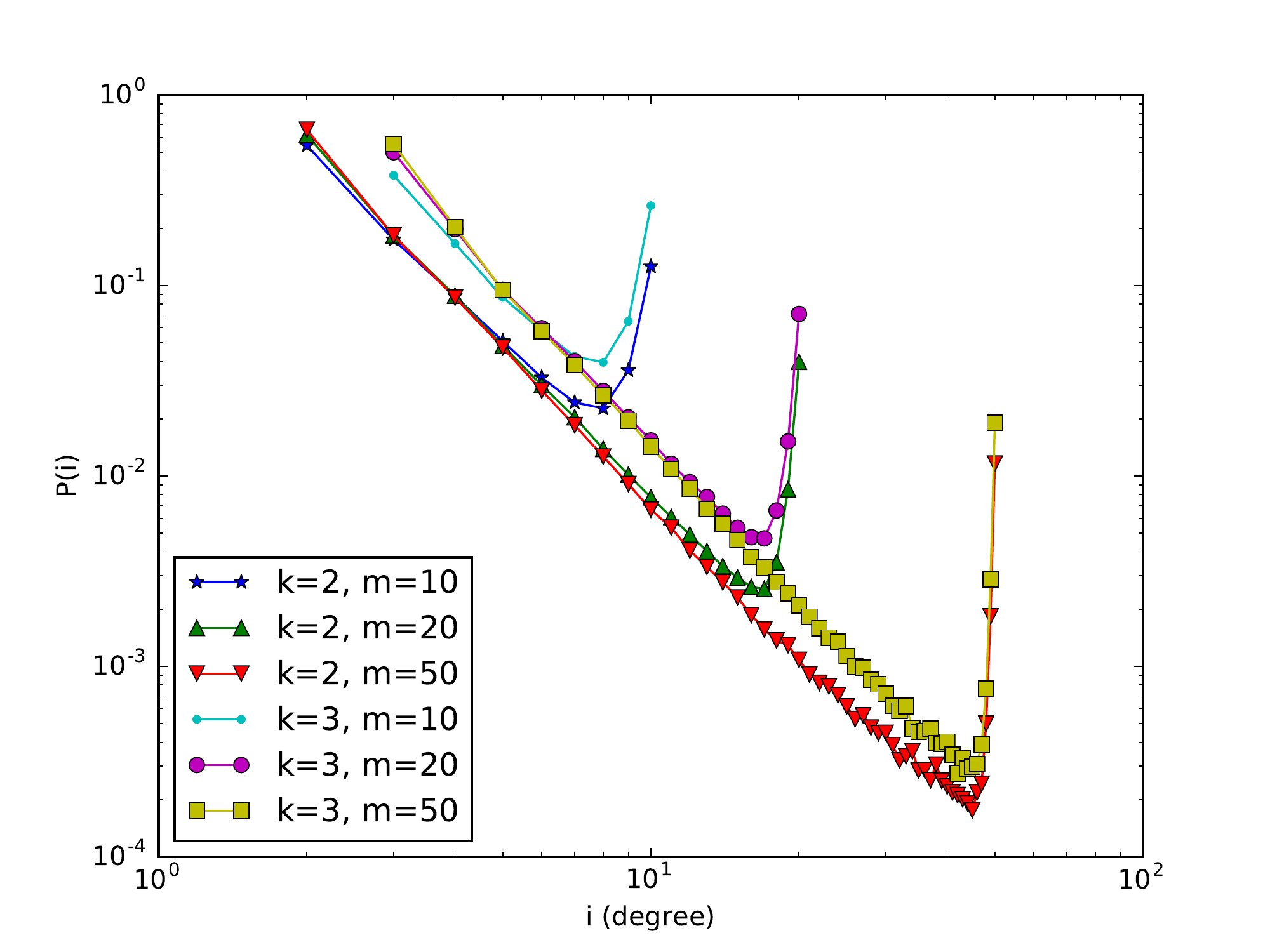}%
\label{subpa_less}}
\caption{
Degree distribution of the topologies where only nodes with degree smaller than $2k$ are randomly removed. $n \approx 50000$, $p=1/3$.}
\label{fig:fig_sim_less}
\end{figure*}

The degree distribution of the topologies where nodes with degree smaller than $2k$ are removed is shown in Figure \ref{fig:fig_sim_less}. The probability to remove a node in each iteration is $p=1/3$. Figure \ref{fig:fig_sim_less} shows all four compared approaches produce the network topologies approximately matching the power-law degree distribution. In theory, a degree distribution following the power-law should be presented as a straight line with slope $-\gamma$ (parallel to the blue dashed line) in the log-log plot. However, it can be observed in Figures \ref{fig:fig_sim_less}.d and \ref{fig:fig_sim_less}.e that the slope of the curve in HAPA and Gaian cases for the range of degrees $[2,4]$ is above the value of $-\gamma$, indicating lower than needed number of low degree nodes for these methods. In contrast, subPA and E-SRA produce the network topology with a perfect matching to the power-law degree distribution at low degrees, undisturbed by removal of nodes with degree smaller than $2k$. 

We estimate the scaling parameter of the produced topologies using the MLE method in \cite{est:method}. In \cite{est:method}, the authors use the Kolmogorov-Smirnov
or KS statistic to quantify the difference between the observed degree distribution and the power-law. The smaller is the KS statistic, the closer is the observed degree distribution to the power-law.

\begin{table}[!t]
\caption{Results of fitness analysis}
\label{table:fitness}
\centering
\setlength\tabcolsep{5pt}
\begin{tabular}{|c|c||c|c|c|c|}
\hline
 Parameters & Method & \begin{tabular}{@{}c@{}}E-SRA \\ ($\gamma = 2.5$)\end{tabular} & HAPA & Gaian & subPA\\
\hline
\hline
 \multirow{2}{*}{$m=20$, $d\geq 6$} & $\gamma$ & 2.505169 & 3.820267 & 3.574570 & 3.885132\\
\hhline{~-----}
 & KS statistic & 0.002982 & 0.102532 & 0.082440 & 0.152262\\
\hline

\multirow{2}{*}{$m=50$, $d\geq 6$} &$\gamma$ & 2.496524 & 2.672919 & 2.667210 & 3.885132\\
\hhline{~-----}
& KS statistic & 0.004415 & 0.117416 & 0.078553 & 0.099203\\
\hline

\multirow{2}{*}{$m=20$, $d< 4$}  & $\gamma$ & 2.502646 & 2.283350 & 2.397095 & 3.967446\\
\hhline{~-----}
& KS statistic & 0.002524 & 0.082762 & 0.023171 & 0.085813\\
\hline

\multirow{2}{*}{$m=50$, $d< 4$} & $\gamma$ & 2.488510 & 2.269304 & 2.387625 & 3.967446\\
\hhline{~-----}
& KS statistic & 0.004728 & 0.075127 & 0.010877 & 0.036065\\
\hline
\end{tabular}
\end{table}

In Table \ref{table:fitness}, the degree of removed nodes is denoted as $d$. When $d\geq6$, only nodes with degree at least 6 are randomly removed from the topology. As seen in Table \ref{table:fitness}, E-SRA produces topologies with a good fit to the scale-free property. The estimated scaling parameters are close to the predefined value $\gamma=2.5$. In the topologies maintained by HAPA and Gaian algorithms, however, the estimated scaling parameters deviate from the scaling parameter, $3.0$ and $2.5$, respectively. In the topologies constructed by subPA in presence of frequent removals, the estimated scaling parameter is larger than $3$. In the cases where removals are not involved, the estimated scaling parameter of subPA is in the range (2,3). In all cases, the KS statistics of HAPA, Gaian and subPA are larger than those of E-SRA.

\subsection{Search Efficiency}

We evaluate the search efficiency over the topologies produced by E-SRA, HAPA, Gaian and subPA algorithms. We consider two search algorithms commonly used in unstructured P2P networks: 1) Flooding (FL), in which every node forwards a query to all the neighbors until the query hits the target. 2) Normalized Flooding (NF), in which every forwarder randomly chooses $k$ (i.e. the minimum degree) neighbors and sends them the query. Thus, the Normalized Flooding
algorithm sets constraints on the number of the messages forwarded by each node. The FL algorithm delivers a query to the destination faster than the NL algorithm in an unstructured P2P network but incurs higher message cost. The Normalized Flooding algorithm sets constraints on the number of the messages forwarded by each node. Both algorithms set time to live (TTL), which is the maximum number of hops a message can traverse, is set up to limit the lifetime of a query in a network. More sophisticated search algorithms in the power-law graphs have been studied in \cite{search:1}.

\begin{figure*}[!t]
\centering
\subfloat[FL]{\includegraphics[width=2.6in]{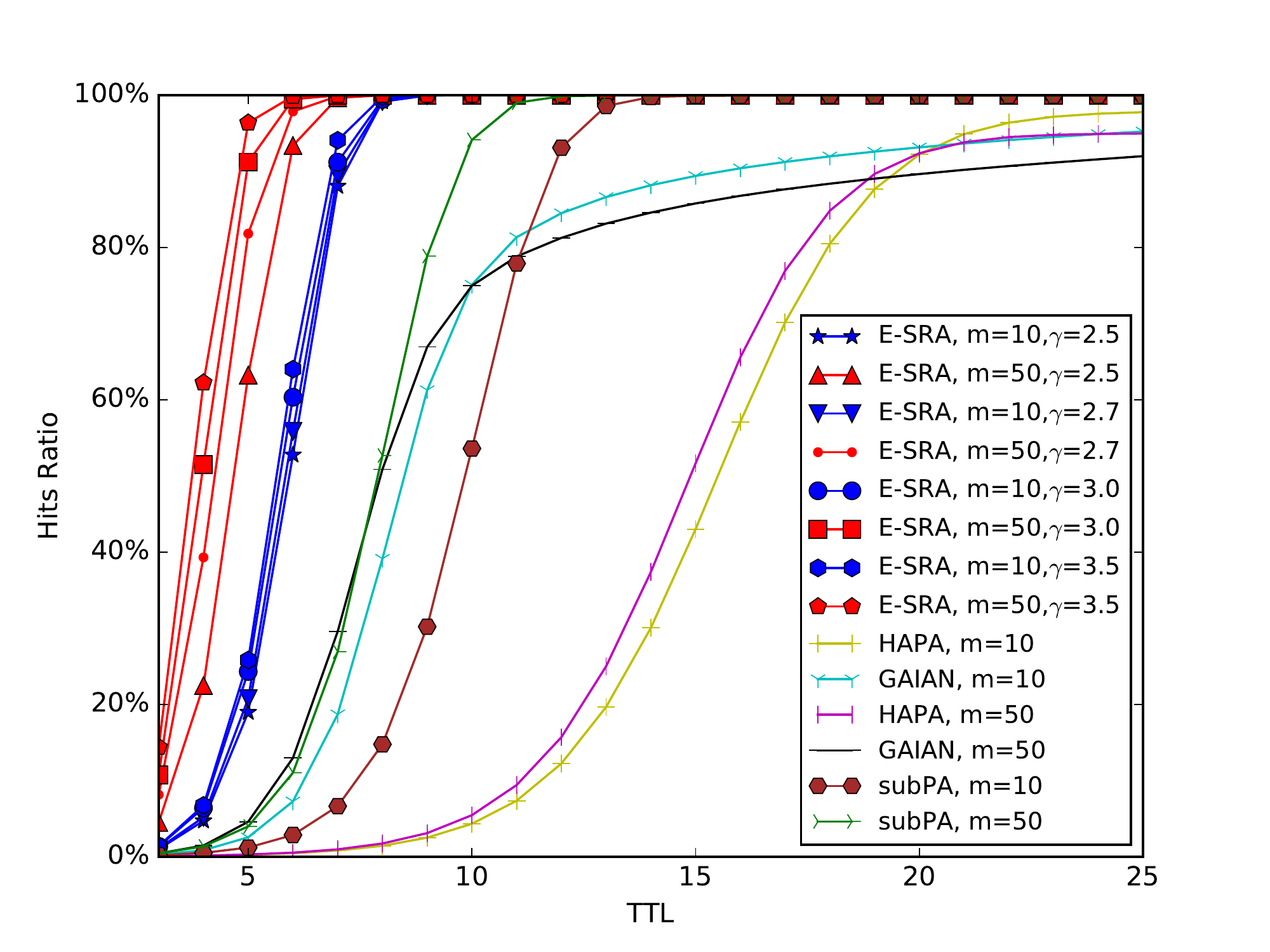}%
\label{removeFL}}
\hfil
\subfloat[NF]{\includegraphics[width=2.6in]{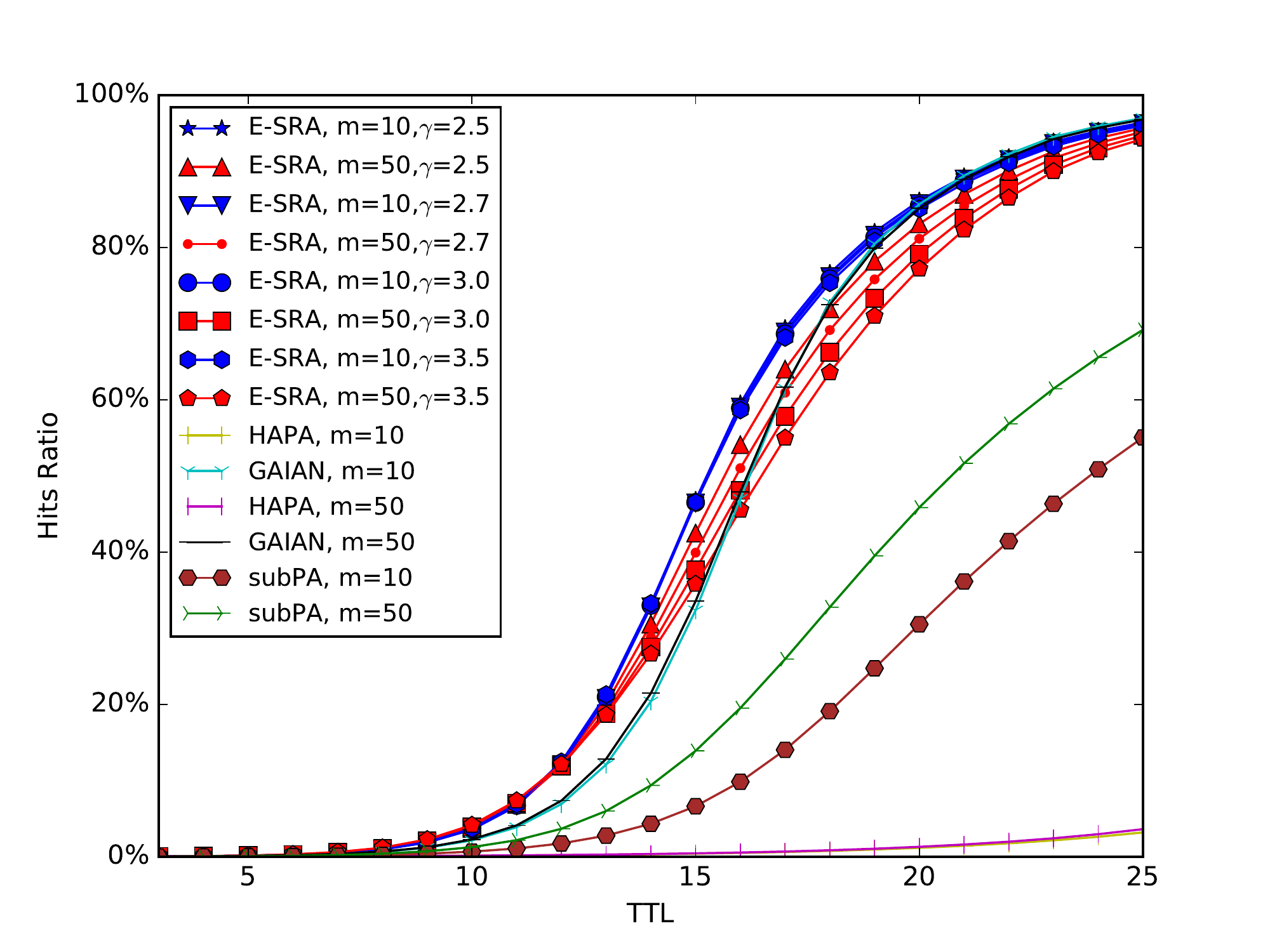}%
\label{removeNF}}
\caption{Search efficiency in networks produced by different approaches and parameters. $n \approx 5 \times 10^4$, $ p=1/3$, $k=2$.}
\label{fig:fig_search}
\end{figure*}

In order to study the performance of the search algorithms, multiple searching processes are simulated on the same topology. With the same parameters, 10 network topologies are constructed. For each topology, 100 nodes are randomly chosen to broadcast a query, whose the average hit ratio is calculated. Since the destinations of the queries are assumed to be uniformly distributed in the network, the expected hit ratio of a query is proportional to the number of nodes it reaches before the TTL expires. In these experiments, the degrees of removed nodes are uniformly distributed in $[k, m]$, and the minimum degree is $k=2$. The results are shown in Figure \ref{fig:fig_search}.

Figure \ref{fig:fig_search}.a shows that a query reaches more than 95\% of the network within 7 hops over the topologies produced by E-SRA with $m=50$ (the red lines); it takes 8 hops for a query to reach 95\% nodes in the topologies with $m=10$ (the blue lines). It shows hit ratio grows fast in the topology constructed by subPA and the query reaches the majority of the network in 10 hops. In the topologies generated by Gaian algorithm, it takes approximately 10 hops for a query to reach 80\% but the spreading process becomes much slower after 12 hops. In HAPA, the hit ratio increases very slowly within the first 10 hops and then increases rapidly. Because the topologies produced by E-SRA highly adhere to the scale-free property, the FL algorithm achieves better performance over these topologies.

The search efficiency of the Normalized Flooding algorithm is shown in Figure \ref{fig:fig_search}.b. One interesting phenomenon is that NF algorithm achieves a higher search efficiency on top of the topologies produced by E-SRA with a small maximum degree $m$. This is because NF only forwards a query to $k$ neighbors, the nodes with high degrees do not have significant advantages over the nodes with low degrees in forwarding queries. In the first $10$ hops, the spreading speed is very slow in the topologies constructed by all four approaches. After 10 hops, the spreading speed becomes much faster. The hit ratio of E-SRA is approximately at least 6\% higher than other approaches with the same TTL in the range $[14, 18]$. And NF algorithm achieves a lower search efficiency in the topologies constructed by subPA than by other models.

 It is worth noting that E-SRA can produce the topologies with the user-defined parameters. Therefore, any proper value of the scaling parameter $\gamma$ could be adopted to produce the best overlay topology according to simulation results. For example, the FL algorithm achieves the best search efficiency on top of the topologies constructed with $m=50$ and $\gamma=3.5$, and the NF algorithm achieves the best search efficiency on top of the topologies constructed with $m=10$, $\gamma=2.5$. E-SRA can use these parameters to construct the desired overlay topologies.
 
 \section{Discussion} 
\label{sec:discuss}
In E-SRA, we assume nodes crash incrementally, one after the other. It guarantees that all neighbors of the node $R$ are alive when $R$ quits and are able to connect to the remaining nodes. However, in realistic situations, a group of connected nodes may crash at the same time. It is possible that $R$ and a neighbor of it fail simultaneously and its neighbor could not PUSH or SHUFFLE correctly as requested by R. In such cases, restoration with global knowledge on the network might be needed. 

It is worth noting that as long as at least one neighbor of the crashed node $R$ is alive, $R$ can be correctly removed from network to preserve the power-law topology. This is because the alive neighbor of $R$ could serve as a coordinator for the crashed neighbors to finish their PUSH/SHUFFLE operations. The nodes, which have the desired degree of those missing PUSH/SHUFFLE operations, need to confirm with this coordinator to finish the corresponding operations. Details of a distributed algorithm for this purpose is beyond the scope of this paper.

In a network where nodes fail with low probability or are guaranteed to recover from failures, a node can only send the \textit{update message}s to its neighbor before it quits. In this way, the message cost is reduced.

It is challenging to preserve the power-law topology while a group of connected nodes fail simultaneously. This is because all the neighbors of a failing node may also crash, thus, the remaining nodes in the network do not know the existence of this failing node nor the degree of it. Consequently, they can not rewire appropriately to preserve the original degree distribution. However, if the knowledge of the nodes about the entire topology increases while the associated cost can be handled, this could be achieved with similar fashion as in our algorithm.

In general, the protocols maintaining topologies with the strict and precise adherence to the power law require higher communication cost than those producing approximate topologies. And simultaneous failures of a group of nodes can make the problem more difficult compared with individual random failures. Thus, some protocols trade such precise adherence to the power law for increased resilience in the face of highly frequent and simultaneous join and leave events. In order to choose appropriate protocols in realistic applications, the trade-off between the benefits of precise and strict adherence to the limited power-law topology and the risk of high communication cost for maintenance should be taken into consideration.

\section{Conclusion} 
\label{sec:conclusion}
 E-SRA, an efficient algorithm for maintaining the limited scale-free topology with dynamic peer participation, is proposed. It produces the overlay topology which improves the P2P network performance. The user can define scaling and cut-off parameters of the overlay network to achieve the best performance. Nodes with any degrees, including hubs, are allowed to be removed from the network freely. Our approach is tolerant to the removal of nodes in any patterns and partially tolerant to node failures by having the neighbors of the failing nodes connecting to the remaining nodes smartly. Simulations have shown that E-SRA outperforms previous growth models by producing overlay topologies with higher adherence to the scale-free property. And search algorithms, including the Flooding algorithm and the Normalized Flooding algorithm, achieve better search efficiency over the topologies produced by E-SRA than by previous growth models. In the future, we plan to study the approach to preserve the power-law distribution under simultaneous failures of a group of nodes. And we are also interested in creating growth models which take user behaviors of P2P networks, such as biased access, into consideration.
 
%\appendices

%\appendix
%\section{Upper Bound Derivation of Message Cost }

\section*{Appendix A. Upper Bound Derivation of Message Cost} 
\label{sec:linearP2Pmsg}
According to the growth model proposed in previous work \cite{scalefree:grow}, when a new node joins the network, it connects to $k$ existing nodes. Among these $k$ nodes, each of the $(1 - \sum_{j=k}^{i} f_j)$ nodes needs to send the \textit{update message}s to $i$ neighbors. Thus, if a node joins, the average number of \textit{update message}s sent is, 
\begin{equation} \label{eq:appendix0}
M_{join} = \sum_{i=k}^{m-1} \big( 1 - \sum_{j=k}^{i} f_j \big) i
\end{equation}
If a node with degree $b$ quits, then $(b-k)$ PUSH operations and $k$ SHUFFLE operations are required. Due to the SHUFFLE operations, the degrees of $(1 - \sum_{j=k}^{i} f_j)$ nodes decrease from $(i+1)$ to $i$. This results in $M_{sh}$ \textit{update message}s.
\begin{equation} \label{eq:appendix2}
M_{sh} = \sum_{i=k}^{m-1} \big( 1 - \sum_{j=k}^{i} f_j \big) i
\end{equation}
The $(b-k)$ PUSH operations change the degrees of $(b-k)$ nodes, which leads to a total of $(\sum_{i=k}^{b-1} i)$ \textit{update message}s. If nodes are randomly removed from the scale-free topology, the average number of \textit{update message}s caused by the PUSH operations is,
\begin{equation} \label{eq:appendix3}
M_{push} =   \sum_{b=k+1}^{m} \big( f_b \sum_{i=k}^{b-1} i \big)
\end{equation}
Since $\sum_{j=k}^{m} f_j = 1$, we have,
\begin{equation}
\sum_{i=k}^{m-1} \big( 1 - \sum_{j=k}^{i} f_j \big) i =\sum_{i=k}^{m-1} \big( \sum_{j=i+1}^{m} f_j \big) i = \sum_{b=k+1}^{m} \Big( f_b \sum_{i=k}^{b-1} i \Big)
\end{equation}
Thus,
\begin{equation} \label{eq:appendix7}
M_{push} = M_{sh} = M_{join} = \sum_{b=k+1}^{m} \big( f_b \sum_{i=k}^{b-1} i \big)
\end{equation}
Using Eq. [\ref{eq:fi2}], the rightmost part of the above equation can be simplified as,
\begin{align} \label{eq:appendix4}
&\sum_{b=k+1}^{m} \Big( f_b \sum_{i=k}^{b-1} i \Big)\\
&= \sum_{b=k+1}^{m-1} \Big( f_b \sum_{i=k}^{b-1} i \Big) + ( 1 - \sum _{b=k}^{m-1} f_b) \sum_{i=k}^{m-1} i\\
&= \sum_{i=k}^{m-1} i + \sum_{b=k}^{m-1} \Big[ f_b \big( \sum_{i=k}^{b-1} i - \sum_{i=k}^{m-1} i \big) \Big]\\
&= \sum_{i=k}^{m-1} i - \sum_{b=k}^{m-1} \Big[ f_b \big( \sum_{i=b}^{m-1} i \big) \Big]
\end{align}
Using the definition of $f_i$ in Eq. [\ref{eq:fi1}], the rightmost part of the above equation can be rewritten as, 
\begin{align} \label{eq:appendix5}
&\sum_{i=k}^{m-1} i - \sum_{b=k}^{m-1} \Big[ \frac{m - 2 k}{ b^\gamma \sum_{j=k}^{m-1} \frac{m-j}{j^\gamma}  } \big( \sum_{i=b}^{m-1} i \big) \Big] \\
&= \frac{(m-1+k)(m-k)}{2} - \frac{m - 2 k}{2} 
\frac{\sum_{b=k}^{m-1} \frac{(m-1+b)(m-b)}{  b^\gamma } } {\sum_{j=k}^{m-1} \frac{m-j}{j^\gamma}} \\
&\leq \frac{(m-1+k)(m-k)}{2} - \frac{m - 2 k}{2} 
\frac{(m-1+k)\sum_{b=k}^{m-1} \frac{(m-b)}{  b^\gamma } } {\sum_{j=k}^{m-1} \frac{m-j}{j^\gamma}} \\
&=  \frac{(m-1+k)(m-k)}{2} - \frac{(m - 2 k)(m - 1 + k)}{2}\\
&= \frac{(m - 1 + k)k}{2} 
\end{align}
Consider a network with $n$ nodes, where $N$ nodes have been added to the network ($N\gg n$). A total of $(N-n)$ nodes have been removed from the network. Every removal results in $(M_{push} + M_{sh})$ \textit{update message}s and every joining results in $M_{join}$ \textit{update message}s. If nodes are randomly added and removed over a long period, the average number of \textit{update message}s sent by a single node is,
\begin{align} \label{eq:appendix6}
 M_{ave} &= \lim_{N\to\infty} \frac{N M_{join} + (N-n)(M_{push} + M_{sh})}{N} \\
 &= 6 \sum_{b=k+1}^{m} \Big( f_b \sum_{i=k}^{b-1} i \Big) \leq 3(m - 1 + k)k
\end{align}
Since the minimum degree $k$ is a constant value, the average number of P2P \textit{update message}s is linear to the hard degree cut-off $m$. Specifically, when $k=2$, the average number of \textit{update message}s sent by a single node is at most $6(m+1)$.

\iffalse
In the worst case, only nodes with the maximum degree $m$ are removed from the network. The $(m-k)$ PUSH operations result in $\sum_{i=k}^{m-1} i = O(m^2)$ \textit{update message}s. The average number of \textit{update message}s sent by a node is $O(m^2)$.
\fi

\iffalse
\section{}
Appendix two text goes here.
\fi

 \section*{Acknowledgement}

%The authors would like to thank...
This work was partially supported by the Army Research Laboratory under Cooperative Agreement Number W911NF-09-2-0053, by the European Commission under the 7th Framework Programme, Grant Agreement Number 316097 and by the Polish National Science Centre, the decision no. DEC-2013/09/B/ST6/02317.

% Can use something like this to put references on a page
% by themselves when using endfloat and the captionsoff option.

\section*{ }

\bibliographystyle{model3a-num-names}

%\section*{References}

%\bibliography{mybibfile}

\end{document}